\documentclass[10pt,a4paper]{article}

\usepackage[T1]{fontenc}
\usepackage[utf8]{inputenc}
\IfFileExists{lmodern.sty}{\usepackage{lmodern}\usepackage{microtype}}{}
\usepackage{amsmath,amssymb,amsthm}
\usepackage{mathtools}
\usepackage[margin=2.5cm]{geometry}
\usepackage{tikz}
\usetikzlibrary{arrows.meta,positioning,bending}
\usepackage{enumitem}
\usepackage{algorithm}
\usepackage[noend]{algpseudocode}
\usepackage[hidelinks]{hyperref}

\theoremstyle{plain}
\newtheorem{theorem}{Theorem}[section]
\newtheorem{proposition}[theorem]{Proposition}
\newtheorem{lemma}[theorem]{Lemma}
\newtheorem{corollary}[theorem]{Corollary}
\theoremstyle{definition}
\newtheorem{definition}[theorem]{Definition}
\newtheorem{example}[theorem]{Example}
\theoremstyle{remark}

\newcommand{\Eh}{\mathbf{W}}
\newcommand{\Hb}{\mathbf{H}}
\newcommand{\Dz}{\mathbf{D}_0}
\newcommand{\Jb}{\mathbf{J}}
\newcommand{\Db}{\mathbf{D}}
\newcommand{\Cb}{\mathbf{C}}
\newcommand{\Sb}{\mathbf{Z}}
\newcommand{\Eb}{\mathbf{E}}

\newcommand{\Rb}{\mathbf{R}}
\newcommand{\Bb}{\mathbf{B}}
\newcommand{\Lb}{\mathbf{L}}
\newcommand{\Ab}{\mathbf{A}}
\newcommand{\Ub}{\mathbf{U}}
\newcommand{\Nb}{\mathbf{N}}
\newcommand{\Tb}{\mathbf{T}}
\newcommand{\Ss}{\mathbf{S}}
\newcommand{\Pp}{\mathcal{P}}
\newcommand{\Oo}{\mathcal{O}}
\newcommand{\Cc}{\mathcal{C}}
\newcommand{\An}{\mathit{An}}
\newcommand{\Pa}{\mathit{Pa}}
\newcommand{\indep}{\perp\!\!\!\perp}
\newcommand{\nindep}{\not\!\perp\!\!\!\perp}
\newcommand{\Proc}[1]{\textup{\textsc{#1}}}
\newcommand{\Fail}{\Proc{fail}}

\algrenewcommand\algorithmicrequire{\textbf{Input:}}
\algrenewcommand\algorithmicensure{\textbf{Output:}}

\title{A fast and sound algorithm to simplify summations in interventional queries expressions}

\author{%
  Ivano Lodato\textsuperscript{1},\quad
  Simran Hiranandani\textsuperscript{1},\quad
  Aditya Varna Iyer\textsuperscript{1,2}\\[0.8em]
  {\small\textsuperscript{1}Allos AI Limited, Hatton Garden, London, EC1 N8LE, UK}\\
  {\small\textsuperscript{2}Clarendon Laboratory, Department of Physics, University of Oxford, Oxford OX1 3PU, UK}\\[0.4em]
  {\small\texttt{ivano.lodato@allos.ai}, \texttt{simran.h@allos.ai},
   \texttt{aditya.iyer@allos.ai}}%
}
\date{}
 
\tikzset{
  vtx/.style={circle,draw,inner sep=1.2pt,minimum size=5.5mm,font=\footnotesize},
  dir/.style={-{Stealth[length=2mm]},thin},
  bid/.style={{Stealth[length=2mm]}-{Stealth[length=2mm]},dashed,thin,gray}
}
 
\begin{document}
\maketitle

\begin{abstract}
\noindent Identification formulas returned by the ID algorithm often contain summations that can
be removed. The procedure proposed in \cite{tikkakarvanen2017} (TK) removes them by combining factors of the
formula into joint distributions, through the chain rule and through the insertion of factors for
missing variables. We make two changes to this
procedure: only the factors that involve the summed variable are combined, and every factor is
written with a minimal conditioning set. These two changes determine the rest of the procedure, in which each
search over a power set is replaced by a single test and the variables to insert are fixed in advance,
and they remove summations that the TK procedure leaves in place. The resulting procedure is sound and, for minimal input terms, complete for each individual elimination with respect to joins and insertions.
\end{abstract}

\section{Introduction}
\label{sec:intro}

A central task of causal inference is to express the effect of an intervention through the
distribution of observed variables. When the causal assumptions are encoded in a diagram $G$, whose
directed edges represent direct causal influence and whose bidirected edges represent latent common
causes, an interventional distribution $P(\mathbf{y}\mid\hat{\mathbf{x}})$ is \emph{identifiable} if it can
be computed from the observational distribution $P(\mathbf{v})$ and $G$ alone
\cite{pearl1995,pearl2009}. Identifiability is characterised by the do-calculus \cite{pearl1995},
whose completeness was shown independently by Shpitser and Pearl \cite{shpitserpearl2006a} and by
Huang and Valtorta \cite{huangvaltorta2006}. The ID algorithm of Shpitser and Pearl
\cite{shpitserpearl2006a} decides it constructively, by a recursion on the \emph{c-components} of
$G$, the maximal sets of vertices connected by bidirected paths, over which the observational
distribution factorises \cite{tianpearl2002,tianpearl2003}, and returns a formula for
$P(\mathbf{y}\mid\hat{\mathbf{x}})$ in the identifiable cases.

The formula is always correct but rarely in its simplest form. It is built from conditional distributions
$P(V\mid\Cb)$, which we call \emph{terms}, with \emph{head} $V$ and \emph{conditioning set} $\Cb$,
combined by products, ratios and nested summations. Simplifying such formulas is not just a matter of elegance: a simpler formula is easier to interpret, since it may reveal a familiar adjustment, and easier to estimate especially when generalized random zeros \cite{Lodato:25} exist in the dataset.\\
An ID interventional formula can be simplified in three (main) ways: common factors in numerator and denominator of a single ratio; cancellations between factors of different ratios; summations that can be worked out. This paper concerns the last problem, in particular how to solve summations in \emph{atomic expressions}
$$\sum_{\Ss}\prod_iP(V_i\mid\Cb_i)$$
which are sums of products of terms. A summation is added to the expression whenever line 2 of ID is recursively called, or when line 4 or one of the exit lines 1 or 6 is called, whether or not it can be carried out.

To show how sums can be simplified, let us consider the query $P(c\mid \hat{b})$ on the simple collider graph $G=a\to c\leftarrow b$ with topological order $a<b<c$. ID returns
\begin{equation}\label{eq:ex-id}
  P(c\mid \hat{b})=\sum_a P(a)\,P(c\mid a,b).
\end{equation}
Since $a$ and $b$ are independent, $P(a)=P(a\mid b)$, and the chain rule combines the two terms into
a joint distribution in which $a$ is a head and nowhere a condition:
\begin{equation}\label{eq:ex-join}
  P(c\mid a,b)\,P(a)=P(a,c\mid b),
  \qquad\text{hence}\qquad
  P(c\mid \hat{b})=\sum_a P(a,c\mid b)=P(c\mid b).
\end{equation}
Every removal of a summation known to us follows this pattern, which we call the \emph{elimination} of the variable: the terms are \emph{assembled} into a joint distribution where the summed variable is a head, the variable is summed out, and the result is
factorised again.

The procedure TK \cite{tikkakarvanen2017} performs these eliminations, by assembling terms in decreasing topological order of their heads by pairwise \emph{joins}
\begin{equation}\label{eq:join-intro}
  P(\Jb\mid\Db)\,P(V\mid\Cb)=P(\Jb,V\mid\Sb),
\end{equation}
applications of the chain rule whose conditioning set $\Sb$ is searched for among the ancestors of
$V$ and the subsets of a set of preceding vertices. When a join fails because the joint carries a
condition $M$ that the next term cannot carry (the second independence condition of INSERT in TK fails), the procedure \emph{inserts} $M$: a term
$P(M\mid\Eb)$ free of the summed variable, with $\Eb$ found by the same kind of search, is
multiplied into the joint and divided out after the summation. The procedure assembles every term
whose head lies between the summed variable and the last head (the summation head), and assumes that every term is
conditioned on (at least) all the ancestors of its head (topological consistency).

Our first change is to assemble only the terms in which $S$ occurs, the \emph{participating}
terms, as one would do on paper, where to carry out a sum over $S$ one looks
at the factors that contain $S$, tries to combine them by the chain rule, supplies a missing factor
where the chain rule requires one, and sums. Our second change is to write every term with a \emph{minimal} conditioning set, from which
no condition can be dropped without changing the term, and this fixes the rest of the procedure. As we will see in Section~\ref{sec:join}, two
minimal terms can be joined at a unique minimal set only, so that a single test replaces the search; the variables to insert and the term inserted for each are fixed in advance
(Section~\ref{sec:insert}).
The resulting procedure, stated in
Section~\ref{sec:algo}, is sound and, with respect to joins and insertions, complete: every expression that it returns is correct, and it eliminates a summation variable whenever joins and insertions can eliminate it (Section~\ref{sec:sc}).
Section~\ref{sec:setting} introduces the setting and the participating terms,
and Section~\ref{sec:concl} concludes.

\section{Setting, minimality and participating terms}
\label{sec:setting}
This section defines the factors used to eliminate a summation variable and the joint distribution they must form.\\
Throughout, $P$ is positive and faithful to $G$, so that conditional independences and
d-separations coincide, and a topological order $\pi$ of the vertices of $G$ is fixed; we write
$u<v$ when $u$ precedes $v$ in $\pi$. A \emph{term} is a factor $P(\mathbf{W}\mid\Cb)$ whose head
set $\mathbf{W}$ is disjoint from its conditioning set $\Cb$; a term with a single head is written
$P(h\mid\Cb)$. Besides the d-separation criterion we use three properties of conditional
independence, valid for every distribution: \emph{symmetry}, by which
$\mathbf{X}\indep\mathbf{Y}\mid\mathbf{Z}$ and $\mathbf{Y}\indep\mathbf{X}\mid\mathbf{Z}$ are
equivalent; \emph{decomposition}, by which
$\mathbf{X}\indep\mathbf{Y}\cup\mathbf{W}\mid\mathbf{Z}$ implies
$\mathbf{X}\indep\mathbf{Y}\mid\mathbf{Z}$; and \emph{weak union}, by which
$\mathbf{X}\indep\mathbf{Y}\cup\mathbf{W}\mid\mathbf{Z}$ implies
$\mathbf{X}\indep\mathbf{Y}\mid\mathbf{Z}\cup\mathbf{W}$.

\begin{definition}\label{def:min}
A condition $c\in\Cb$ is \emph{idle} in $P(\mathbf{W}\mid\Cb)$ when
$P(\mathbf{W}\mid\Cb\setminus\{c\})=P(\mathbf{W}\mid\Cb)$, that is, when
$\mathbf{W}\indep c\mid\Cb\setminus\{c\}$. The conditioning set $\Cb$ is \emph{minimal} when no
condition is idle.
\end{definition}

A minimal conditioning set is obtained from any conditioning set by dropping idle conditions one
at a time until none is left: this is the function \Proc{Minimize} of Section~\ref{sec:algo}.
Minimality is a property of the term and is not determined by the parents or ancestors of its head:
a minimal conditioning set need not contain the ancestors of the head, contrary to the
topological consistency assumed in \cite{tikkakarvanen2017}.\footnote{In \cite{tikkakarvanen2017} an
atomic expression is called \emph{topologically consistent} when $\An(V_i)_G\subseteq\Cb_i\subseteq
V_i^{\pi}$ for every term, $V_i^{\pi}$ being the set of predecessors of $V_i$
\cite[Definition~4]{tikkakarvanen2017}, and ID outputs are stated to have this property, every
term being written as $P(v_i\mid v_i^{\pi})$ \cite[Appendix~A]{tikkakarvanen2017}. The value of such a term
depends only on $C_i=\Pa(T_i)\setminus\{V_i\}$ (see below), for which the first inclusion fails in
general: in the chain $a\to b\to c$, $C_c=\{b\}$ while $\An(c)=\{a,b\}$.} The uniqueness result of
Section~\ref{sec:join} holds for minimal terms only.

Minimality is already imposed by the ID algorithm, as all the terms returned by it are minimal \cite[Lemma~1]{tianpearl2003}: to see that, let $V_i$ be a vertex and $T_i$ the c-component of $V_i$ in the subgraph induced by $V_i$ and its $\pi$-predecessors, and let
$C_i$ consist of the other members of $T_i$ and the parents of all members of $T_i$. By the
c-component factorisation \cite{tianpearl2003}, the conditionals $P(v_i\mid v_i^{\pi})$ formed at
lines~6 and~7 of ID \cite{shpitserpearl2006a} equal $P(V_i\mid C_i)$. No condition $c\in C_i$ is
idle, because $c$ either belongs to $T_i$ or is a parent of some $t\in T_i$, and therefore reaches $V_i$
through a bidirected path inside $T_i$, preceded in the second case by the edge $c\to t$. Every
intermediate vertex of this path lies in $T_i$, hence is conditioned, and is a collider, so that the
path is open given $C_i\setminus\{c\}$ and, under faithfulness, $V_i\nindep c\mid C_i\setminus\{c\}$.
Since $T_i$ lies among the predecessors of
$V_i$, these terms are also conditioned only on variables below their heads. The procedure
maintains this property (Section~\ref{sec:insert}), and we assume it throughout.

The simplification uses four rewritings:
\begin{enumerate}[label=(R\arabic*),leftmargin=*]
  \item A conditional distribution sums to one over its head: $\sum_v P(v\mid\Cb)=1$ whenever
    $v\notin\Cb$.
  \item A factor in which the summed variable does not occur can be taken out of the sum:
    $\sum_U(EF)=E\sum_UF$ when $U$ does not occur in $E$.
  \item Products and sums may be reordered and regrouped, by commutativity and associativity.
  \item A summation over heads of a joint distribution is a marginalisation: if a product of
    terms equals $P(\Eh\mid\Dz)$ and $U\subseteq\Eh$, its sum over $U$ is $P(\Eh\setminus
    U\mid\Dz)$, which the chain rule writes again as a product of terms; for instance
    $\sum_{A,C}P(A)P(B\mid A)P(C\mid A,B)=P(B)$.
\end{enumerate}
The first three apply to any product of terms, while (R4) applies only to terms that together
form a joint distribution.
To \emph{eliminate}
a summation variable $S$ means to assemble terms into a joint distribution $P(\Jb\mid\Db)$ with
$S\in\Jb$ and $S\notin\Db$, to sum over $S$, and to factorise the result. The factorisation writes
each surviving term with all the variables that the chain rule assigns to it, a set that contains
idle conditions in general: it is at this point that minimality must be restored.

We recall TK in the form needed for comparison
\cite[Algorithms 1--3]{tikkakarvanen2017}. Let $\sum_{\Ss}\prod_iP(V_i\mid\Cb_i)$ be atomic, $W$
the set of its variables, and $S\in\Ss$. Terms in which no summation variable occurs are first
extracted \cite[Algorithm~5]{tikkakarvanen2017}. To eliminate $S$, \Proc{Simplify} assembles, in
decreasing order of their heads, \emph{every} remaining term whose head lies between $S$ and the
$\pi$-last head; the variables of $W$ in that range that head no term are the \emph{missing}
variables. \Proc{Join} combines the current joint $P(\Jb\mid\Db)$ with the next term
$P(V\mid\Cb)$ by trying, for each subset $\mathbf{P}$ of $\Jb^\pi\setminus\An^*(V)$, the
conditioning set $\An(V)\cup\mathbf{P}$;\footnote{The text of \cite{tikkakarvanen2017} defines
$\Jb^\pi$ as the vertices preceding every vertex of $\Jb$, while its worked example
\cite[Section~5]{tikkakarvanen2017} uses the vertices preceding the $\pi$-last vertex of $\Jb$,
other than those of $\Jb$; we adopt the latter, which is the more favourable to the procedure.
$\An^*(V)$ denotes $\An(V)\cup\{V\}$, and ancestors are taken among the variables of $W$.} a candidate is accepted when
$\Jb\indep\mathbf{Y}_1\mid\Db\setminus\mathbf{Y}_1$ and $V\indep\mathbf{Y}_2\mid\Cb\setminus\mathbf{Y}_2$,
with $\mathbf{Y}_1=(\An^*(V)\cup\mathbf{P})\triangle\Db$ and
$\mathbf{Y}_2=(\An(V)\cup\mathbf{P})\triangle\Cb$. When no candidate is accepted, \Proc{Insert}
tries a missing variable $M\in\Db\setminus\Cb$ with the conditioning sets $\An(M)\cup\mathbf{P}$,
accepting the first for which the joint absorbs the new term and $M\indep S$ given it. The
summation is then carried out on the joint and the inserted terms are divided out; no
minimality check is performed.

The first change we make concerns the first step, the choice of the terms to be assembled. Let
$\sum_{\Ss}\prod_{t\in\Tb}t$ be an atomic expression: $\Ss$ is its set of summation variables, and
$\Tb$ is its set of terms, each a conditional distribution $t=P(V\mid\Cb)$, given by its head $V$
and its conditioning set $\Cb$. Let $S\in\Ss$ be the variable to be eliminated. Like every summation
variable in an output of ID, $S$ is the head of one term of the expression: the output is a
probabilistic expression, in which a summation marginalises a distribution over some of its own
variables, and a variable that heads no term cannot be summed.

\begin{definition}\label{def:part}
The \emph{participating} terms $\Pp_S\subseteq\Tb$ are those in which $S$ occurs, as head or as
condition; the others form $\Oo_S=\Tb\setminus\Pp_S$. We write $\Hb$ for the heads of $\Pp_S$,
$C_h$ for the conditioning set of the term headed by $h\in\Hb$, and
\[
  \Nb:=\Bigl(\bigcup_{h\in\Hb}C_h\Bigr)\setminus\Hb
\]
for the variables on which participating terms are conditioned without heading one of them.
\end{definition}

By (R2), $\sum_S\prod_{t\in\Tb}t=\prod_{t\in\Oo_S}t\cdot\sum_S\prod_{t\in\Pp_S}t$, so only
$\Pp_S$ has to be assembled into a joint distribution. TK
assembles more: every term whose head lies between $S$ and the last head of the expression, together
with terms inserted for missing variables. The end result of both procedures, if successful, is an \emph{assembly}.

\begin{definition}\label{def:assembly}
An \emph{assembly for $S$} is a joint distribution $P(\Ab\mid\Bb)$, with $\Hb\subseteq\Ab$ and $S\in\Ab$,
whose heads are taken in decreasing topological order $a_r>\dots>a_1$, and which is written by the
chain rule as
\[
  P(\Ab\mid\Bb)=\prod_{i=1}^{r}P\bigl(a_i\mid\Eb_{a_i}\bigr),\qquad
  \Eb_{a_i}:=\Bb\cup\{a_1,\dots,a_{i-1}\},
\]
where the factor of each participating head $h$ is its term, $P(h\mid C_h)=P(h\mid\Eb_h)$, and the
factor of every other head is a further term that does not depend on $S$, either a term of $\Oo_S$ or a term to be inserted.
\end{definition}
The summation over $S$ is then a marginalisation of the joint, after which the inserted terms are
divided out. Since every condition of a participating term is a variable of the joint,
$\Nb\subseteq\Ab\cup\Bb$.

\begin{definition}\label{def:fullreduced}
For $\Rb\subseteq\Nb$, let $\Lb:=\Nb\setminus\Rb$. A \emph{reduced} assembly is an assembly
$P(\Hb\cup\Rb\mid\Lb)$: its heads are the participating heads and the variables of $\Rb$, its
conditions are the other variables of $\Nb$, and its further terms are only inserted terms for the
variables of $\Rb$. A \emph{full} assembly $P(\Ab\mid\Bb)$ may contain, besides the participating
terms, terms of $\Oo_S$, called \emph{retained} terms, and inserted terms for missing variables,
so that $\Ab$ and $\Bb$ may contain variables outside $\Hb\cup\Nb$. The \emph{reduction} of a full
assembly $P(\Ab\mid\Bb)$ is the reduced assembly with $\Rb:=\Ab\cap\Nb$, so that $\Lb=\Bb\cap\Nb$:
it keeps the participating heads and the variables of $\Rb$ in the same order, conditions them on
$\Lb$, and drops the variables of $\Ub:=(\Ab\cup\Bb)\setminus(\Hb\cup\Nb)$, namely the heads of
retained terms, the missing variables outside $\Hb\cup\Nb$, and the conditions of the joint outside
$\Nb$.
\end{definition}

TK produces full assemblies, and the procedure of
Section~\ref{sec:algo} reduced ones; the following proposition relates the two.

\begin{proposition}\label{prop:reduce}
If a full assembly for $S$ exists, its reduction is an assembly for $S$, in the same order of
assembly.
\end{proposition}

By the proposition, whatever a full assembly eliminates is eliminated by the participating terms
alone, with inserted terms for the variables of $\Rb=\Ab\cap\Nb$. Only variables of $\Nb$ therefore
need to be inserted. Such a variable may already exist in the expression and head a term of $\Oo_S$; in our procedure that term then remains outside the
summation, and its head is inserted with the term that the chain rule requires
(Example~\ref{ex:frontdoor}). Definition~\ref{def:assembly} takes the heads in decreasing
topological order, as do Definition~5 of \cite{tikkakarvanen2017} and the procedure of
Section~\ref{sec:algo}, and this choice is not a restriction, because every term depends only on variables below
its head, so that a joint assembled in another order can be rearranged into this one without changing its
terms (Appendix~\ref{app:step2}).

The converse of the proposition fails, because a retained term imposes two constraints that a
reduced assembly avoids. First, it must \emph{fit}, that is, it must be the factor that the chain
rule assigns to its head in the full joint. Second, it \emph{adds conditions}: its conditions enter
the joint, and every participating term must then be independent of those it does not carry. In
Example~\ref{ex:setaside} below the added condition $X$ makes the join with the participating term $P(B\mid A)$
fail, and in Example~\ref{ex:frontdoor} of Section~\ref{sec:insert} the retained term does not fit:
in both, a reduced assembly eliminates a variable that TK cannot eliminate. These comparisons
concern one summation at a time, whereas the outcome of a whole run also depends on the order in
which the summation variables are tried. In Example~\ref{ex:order} of the appendix, with the
summation variables tried in a fixed order, the procedure of this paper removes fewer summations than TK,
because it eliminates a variable that TK cannot and this blocks later eliminations; tried in
another order, it reaches the same result as TK.

\begin{example}[a retained term above the participating terms]\label{ex:setaside}
Consider the graph below, with the topological order $X<A<B<Y$, and the query $P(Y\mid \hat{X})$.
\begin{center}
\begin{tikzpicture}[node distance=14mm]
\node[vtx] (A) {$A$};
\node[vtx,right=of A] (B) {$B$};
\node[vtx,right=of B] (Y) {$Y$};
\node[vtx,above=9mm of B] (X) {$X$};
\draw[dir] (A) -- (B);
\draw[dir] (B) -- (Y);
\draw[dir] (X) -- (Y);
\draw[bid] (X) to[bend right=35] (B);
\end{tikzpicture}
\end{center}
ID returns
\[
  P(Y\mid \hat{X})=\sum_{A,B}P(A)\,P(B\mid A)\,P(Y\mid X,B).
\]
The summation variables are tried in decreasing order, beginning with $B$, whose participating terms are
$P(Y\mid X,B)$ and $P(B\mid A)$. Their join requires $B\indep X\mid A$, which fails because of
$X\leftrightarrow B$; no variable is inserted, since the only condition of the joint that
$P(B\mid A)$ does not carry is $X<B$. The participating terms of $A$ are $P(A)$ and $P(B\mid A)$
only, while $P(Y\mid X,B)$ remains outside the summation. By the chain rule
$P(A)P(B\mid A)=P(A,B)$, and summing over $A$ gives $P(B)$, so that
\[
  P(Y\mid \hat{X})=\sum_B P(B)\,P(Y\mid X,B),
\]
the back-door formula. A second attempt on $B$ requires $P(B)=P(B\mid X)$, which fails again; the
remaining summation is the adjustment for $B$, and no rewriting removes it.

To eliminate $A$, TK must assemble the heads $Y$, $B$ and $A$, beginning with the join of
$P(Y\mid X,B)$ and $P(B\mid A)$. Here $\An(B)=\{A\}$ and $\Jb^\pi\setminus\An^*(B)=\{X\}$, so the
candidate sets are $\{A\}$, which fails $Y\indep\{A,X\}\mid B$ because of $X\to Y$, and
$\{A,X\}$, which passes $Y\indep A\mid X,B$ but fails $B\indep X\mid A$ because of
$X\leftrightarrow B$. No variable is missing between $A$ and $Y$, so the elimination fails; the
elimination of $B$ begins with the same join and fails as well, and the expression is left
unchanged. The same happens in each of the three topological orders of the graph; in those where
$X$ follows $A$, $X$ is missing and its insertion fails too. The failure is caused by the term
$P(Y\mid X,B)$, which is retained although it does not contain $A$, and whose join with
$P(B\mid A)$ is impossible at any set.
\end{example}

Participation is read off the conditioning sets as written. A term that carries an idle condition
on $S$ therefore takes part in the elimination of $S$, although it does not depend on $S$, and
must be assembled with the other participating terms; if that assembly is impossible, the
elimination is lost. Such conditions do not occur in the terms emitted by ID, but they occur after
a summation unless minimality is restored (Example~\ref{ex:restore}), and they are written by the search of
\cite{tikkakarvanen2017}, whose joints are conditioned on all the ancestors of the joined head
(Example~\ref{ex:order}). In the procedure of Section~\ref{sec:algo}, every term written after a
summation is minimised.

\section{The conditioning set of a join}
\label{sec:join}

The first step in building a joint distribution in which $S$ is a head is to join terms, one at a
time, by the chain rule.
The highest term forms the initial joint, and each following term $P(V\mid\Cb)$ is joined to the
joint $P(\Jb\mid\Db)$ assembled so far. A join must choose the conditioning set of the new joint:
TK searches for it among the sets $\An(V)\cup\mathbf{P}$, whereas for minimal terms there is only
one candidate, a fixed conditioning set at which two independences decide the join.

Since every term is conditioned only on variables below its head, and the heads of the joint lie
above $V$, the set $\Jb$ is disjoint from $\Cb\cup\{V\}$. We write
\[
  \Cc:=(\Db\cup\Cb)\setminus\{V\},\qquad\text{so that}\qquad
  \Cb\subseteq\Cc,\quad\Db\subseteq\Cc\cup\{V\}.
\]

\begin{theorem}[The unique join set]\label{thm:join}
Let $P(\Jb\mid\Db)$ and $P(V\mid\Cb)$ be minimal.
\begin{enumerate}[label=(\roman*),nosep]
  \item The two terms are joined at some set $\Sb$, that is,
    $P(\Jb\mid\Db)\,P(V\mid\Cb)=P(\Jb,V\mid\Sb)$, if and only if they are joined at $\Cc$,
    \begin{equation}\label{eq:joinU}
      P(\Jb\mid\Db)\,P(V\mid\Cb)=P(\Jb,V\mid\Cc),
    \end{equation}
    and this holds if and only if
    \begin{equation}\label{eq:testgraph}
      \Jb\indep(\Cb\cup\{V\})\setminus\Db\,\big|\,\Db
      \qquad\text{and}\qquad
      V\indep(\Db\setminus\{V\})\setminus\Cb\,\big|\,\Cb.
    \end{equation}
    In that case $\Cc\subseteq\Sb$, and every condition in $\Sb\setminus\Cc$ is idle in
    $P(\Jb,V\mid\Sb)$.
  \item If \eqref{eq:joinU} holds, $\Cc$ is a minimal conditioning set for $P(\Jb,V\mid\Cc)$.
\end{enumerate}
\end{theorem}

The two independences in \eqref{eq:testgraph} state that $P(\Jb\mid\Db)=P(\Jb\mid\Cc,V)$ and
$P(V\mid\Cb)=P(V\mid\Cc)$: each term can be written with the conditioning set that the chain rule
assigns to it in the joint $P(\Jb,V\mid\Cc)$.

\begin{proof}
(i) The statement contains three assertions: the terms are joined at some set, they are joined at
$\Cc$, and the independences \eqref{eq:testgraph} hold. We prove that the third implies the second,
the second the first, and the first the third, so that the three are equivalent and both
equivalences of (i) hold. If \eqref{eq:testgraph} holds, the chain rule gives
$P(\Jb\mid\Db)P(V\mid\Cb)=P(\Jb\mid\Cc,V)\,P(V\mid\Cc)=P(\Jb,V\mid\Cc)$, which is
\eqref{eq:joinU}. If \eqref{eq:joinU} holds, the terms are joined at a set, namely $\Sb=\Cc$.
Finally, suppose that
$P(\Jb\mid\Db)P(V\mid\Cb)=P(\Jb,V\mid\Sb)$ for some $\Sb$. Summing both sides over $\Jb$, which
does not occur in $P(V\mid\Cb)$, removes the first factor by (R1) and marginalises the
right-hand side, leaving
\begin{equation}\label{eq:V}
  P(V\mid\Cb)=P(V\mid\Sb);
\end{equation}
dividing the join identity by \eqref{eq:V}, which positivity permits, leaves
\begin{equation}\label{eq:Jj}
  P(\Jb\mid\Db)=P(\Jb\mid V,\Sb).
\end{equation}
If some $c\in\Cb$ were not in $\Sb$, the right-hand side of \eqref{eq:V} would not depend on $c$,
so that $c$ would be idle in $P(V\mid\Cb)$, against minimality; hence $\Cb\subseteq\Sb$. The same
argument applied to \eqref{eq:Jj} gives $\Db\subseteq\Sb\cup\{V\}$, and together these yield
$\Cc\subseteq\Sb$. Equation~\eqref{eq:V} states $V\indep\Sb\setminus\Cb\mid\Cb$; since
$\Sb\setminus\Cb$ is the union of $\Cc\setminus\Cb$ and $\Sb\setminus\Cc$, decomposition gives
$V\indep\Cc\setminus\Cb\mid\Cb$, which is the second independence of \eqref{eq:testgraph}
because $(\Db\setminus\{V\})\setminus\Cb=\Cc\setminus\Cb$. Equation~\eqref{eq:Jj} states
$\Jb\indep(\Sb\cup\{V\})\setminus\Db\mid\Db$; since $(\Cb\cup\{V\})\setminus\Db$ is contained in
$(\Sb\cup\{V\})\setminus\Db$, decomposition gives the first independence of
\eqref{eq:testgraph}, which closes the cycle. Moreover \eqref{eq:joinU} then holds, and since both
$P(\Jb,V\mid\Sb)$ and
$P(\Jb,V\mid\Cc)$ equal the product of the two terms, the conditions in $\Sb\setminus\Cc$ are
idle.

(ii) If a condition $c\in\Cc$ were idle in $P(\Jb,V\mid\Cc)$, the two terms would be joined at
$\Sb=\Cc\setminus\{c\}$, and (i) would give $\Cc\subseteq\Cc\setminus\{c\}$, which is
impossible.
\end{proof}

By (ii), a join of minimal terms at $\Cc$ produces a minimal joint, so that along a sequence of
joins every joint is minimal, the theorem applies at every step, and no minimality check is
needed between joins. The theorem also reduces the search of \cite{tikkakarvanen2017} to a single
set. Suppose the search joins $P(\Jb\mid\Db')$ with $P(V\mid\Cb)$ at some set
$\Sb=\An(V)\cup\mathbf{P}$, where $\Db'$ may contain, besides a minimal set $\Db$, conditions that
earlier joins have written although they are idle. The two products have the same value, so the
minimal pair is joined at $\Sb$ as well, and (i) gives $\Cc\subseteq\Sb$ and the join at $\Cc$. The test at $\Cc$ therefore succeeds whenever the search does, and it is a single test, whereas the search examines up to $2^{|\Jb^\pi\setminus\An^*(V)|}$ candidate sets before finding that the join is
possible or impossible (Example~\ref{ex:cost}). The single test is conclusive for minimal terms only, as we shall see in Example~\ref{ex:restore} below.

\begin{example}[one test instead of a search]\label{ex:cost}
Consider the graph below, with the topological order $X<Z<W<A<B<Y$, and the query
$P(W,Y\mid \hat{X},\hat{Z})$.
\begin{center}
\begin{tikzpicture}[x=1.6cm, y=0.9cm]
  \node[vtx] (X) at (0,1.5) {$X$}; \node[vtx] (A) at (0,0.5) {$A$};
  \node[vtx] (Z) at (-1,-0.5) {$Z$}; \node[vtx] (W) at (0,-0.5) {$W$}; \node[vtx] (B) at (0,-1.5) {$B$};
  \node[vtx] (Y) at (2,0) {$Y$};
  \draw[dir] (Z)--(W); \draw[dir] (X)--(Y); \draw[dir] (A)--(Y); \draw[dir] (W)--(Y); \draw[dir] (B)--(Y);
  \draw[bid] (X) to[bend right=50] (A);
\end{tikzpicture}
\end{center}
ID returns
\[
  P(W,Y\mid \hat{X},\hat{Z})=\sum_{A,B}P(W\mid Z)\,P(A)\,P(B)\,P(Y\mid X,W,A,B).
\]
The participating terms of $B$ are $P(Y\mid X,W,A,B)$ and $P(B)$. By Theorem~\ref{thm:join} their
join is fixed at $\Cc=(\{X,W,A,B\}\cup\emptyset)\setminus\{B\}=\{X,W,A\}$; the first independence
of \eqref{eq:testgraph} is vacuous, and the second, $B\indep\{X,W,A\}$, holds, since $B\to Y$ is the
only edge at $B$. The joint is $P(Y,B\mid X,W,A)$, and summing over $B$ gives $P(Y\mid X,W,A)$,
which is minimal. The participating terms of $A$ are then $P(Y\mid X,W,A)$ and $P(A)$; their join
is fixed at $\Cc=\{X,W\}$, and its second independence, $A\indep\{X,W\}$, fails because of
$X\leftrightarrow A$. By the theorem no set joins the two terms, and the procedure stops at
\[
  P(W,Y\mid \hat{X},\hat{Z})=\sum_{A}P(W\mid Z)\,P(A)\,P(Y\mid X,W,A)
\]
after two tests.
 
The procedure of TK reaches the same expression after $22$ candidate sets, $14$ for $B$ and $8$ for $A$. For $B$ it joins the same two terms;
$\An(B)=\emptyset$ and $\Jb^\pi\setminus\An^*(B)=\{X,Z,W,A\}$, so the candidates are the $16$
subsets of $\{X,Z,W,A\}$, which we enumerate from the smallest. Every set that omits $X$, $W$ or $A$ fails,
because $Y$ depends on the omitted variable, and $\{X,W,A\}$, the fourteenth, is accepted: it is
$\Cc$. For $A$ the candidates are the $8$ subsets of $\{X,Z,W\}$. Those that omit $X$ or $W$ fail
on the dependence of $Y$, and $\{X,W\}$ and $\{X,Z,W\}$ fail because $A\nindep X$; all eight must
be tried before the join is abandoned, and no variable is missing between $A$ and $Y$. Every set
that the two searches reject is excluded beforehand by Theorem~\ref{thm:join}.
 
The difference grows with the graph, as the following family shows. Add $m$ intervened variables $X_1,\dots,X_m$, placed after
$X$ in the order, each with the single edge $X_i\to Y$, and consider
$P(W,Y\mid \hat{X},\hat{X}_1,\dots,\hat{X}_m,\hat{Z})$. ID returns the same expression with $X_1,\dots,X_m$ among the
conditions of the term for $Y$, and the procedure of this paper still performs two tests, at
$\Cc=\{X,X_1,\dots,X_m,W,A\}$ and at $\Cc=\{X,X_1,\dots,X_m,W\}$. For TK the join of $B$ now has
$2^{m+4}$ candidate sets and is accepted, at $\Cc$, only after every smaller set ($2^{m+4}-m-3$ sets) has failed; the
join of $A$ is abandoned only after all the $2^{m+3}$ subsets of $\{X,X_1,\dots,X_m,Z,W\}$ have
failed.
\end{example}
\begin{example}[the fixed join needs minimal terms]\label{ex:restore}
Consider the graph below, with the topological order $A<B<X<Z<V<W<Y$, and the query
$P(Z,Y\mid \hat{A},\hat{B},\hat{X},\hat{V},\hat{W})$.
\begin{center}
\begin{tikzpicture}[x=1.7cm, y=1.2cm]
  \node[vtx] (A) at (0,0) {$A$}; \node[vtx] (B) at (1,0) {$B$}; \node[vtx] (X) at (2,0) {$X$};
  \node[vtx] (W) at (3,0) {$W$}; \node[vtx] (Y) at (4,0) {$Y$};
  \node[vtx] (Z) at (2,1) {$Z$}; \node[vtx] (V) at (3,-1) {$V$};
  \draw[dir] (A)--(B); \draw[dir] (B)--(X); \draw[dir] (X)--(W); \draw[dir] (W)--(Y); \draw[dir] (V)--(Y);
  \draw[bid] (A) to[bend left=15] (Z);
  \draw[bid] (X)--(Z);
  \draw[bid] (Z) to[bend left=15] (Y);
  \draw[bid] (X)--(V);
\end{tikzpicture}
\end{center}
ID returns
\begin{gather*}
  P(Z,Y\mid \hat{A},\hat{B},\hat{X},\hat{V},\hat{W})=
  \frac{\sum_{A,X,V,Y}Q}{\sum_{A,X,Z,V,Y}Q}\;\frac{\sum_{A,X}Q}{\sum_{A,X,Y}Q},\\
  Q=P(A)\,P(X\mid B)\,P(Z\mid A,B,X)\,P(V\mid A,B,X,Z)\,P(Y\mid A,B,X,Z,V,W),
\end{gather*}
where the summations of the fractions bind their own variables. The outcomes $Z$ and $Y$ form the
only c-component of the graph without the intervened variables, and their c-component in $G$ is
$\{A,X,Z,V,Y\}$, so that ID reaches line~7 at once, with the product $Q$ of the five terms of this
c-component, written with the minimal conditioning sets of Section~\ref{sec:setting}. In the subgraph on the c-component, $A$ and
$X$ are not ancestors of $Z$ or $Y$ and are summed at line~2, and line~6 writes the distribution of
$Z$, and that of $Y$ given $Z$ and $V$, as the two ratios. The first denominator sums every term
to one, so that the first fraction, the distribution of $Z$, is its numerator,
\[
  \sum_{A,X,V,Y}P(A)\,P(X\mid B)\,P(Z\mid A,B,X)\,P(V\mid A,B,X,Z)\,P(Y\mid A,B,X,Z,V,W),
\]
an atomic expression with minimal terms, whose summation variables are tried in decreasing order.
 
The term for $Y$ is the only one that contains $Y$, and it sums to one; the same then holds for
$V$. This leaves $\sum_{A,X}P(A)\,P(X\mid B)\,P(Z\mid A,B,X)$. The participating terms of $X$ are
$P(Z\mid A,B,X)$ and $P(X\mid B)$. By Theorem~\ref{thm:join} their join is fixed at
$\Cc=(\{A,B,X\}\cup\{B\})\setminus\{X\}=\{A,B\}$; the first independence of \eqref{eq:testgraph} is
vacuous, since $\{B,X\}\subseteq\{A,B,X\}$, and the second, $X\indep A\mid B$, holds: the path
$X\leftarrow B\leftarrow A$ is blocked at $B$, and every other path at a collider, $Z$ or $Y$, that
is not given. The joint is $P(Z,X\mid A,B)$, and summing over $X$ gives $P(Z\mid A,B)$.
 
This term is not minimal: $Z\indep B\mid A$, because $B\leftarrow A\leftrightarrow Z$ is blocked at
$A$, and every path through $X$ at a collider, $X$ or $Y$, that is not conditioned on. The condition $A$
cannot be dropped, because of $A\leftrightarrow Z$, so the minimal term is $P(Z\mid A)$. With it,
the participating terms of $A$ are $P(Z\mid A)$ and $P(A)$, their join at $\emptyset$ involves only
vacuous tests, and summing over $A$ gives $P(Z)$: the four summations are removed, and the first
fraction is $P(Z)$.
 
Suppose instead that the term is kept as the chain rule writes it, $P(Z\mid A,B)$, with the idle
condition $B$ inherited from $P(X\mid B)$. The participating terms of $A$ are $P(Z\mid A,B)$ and
$P(A)$, and the set of Theorem~\ref{thm:join} is $\Cc=(\{A,B\}\cup\emptyset)\setminus\{A\}=\{B\}$.
The first independence of \eqref{eq:testgraph} is vacuous, and the second, $A\indep B$, fails
because of $A\to B$; the insertion of $B$, which lies above $A$, requires the same independence
(Section~\ref{sec:insert}). The two terms can nevertheless be joined, since
$P(Z\mid A,B)\,P(A)=P(Z\mid A)\,P(A)=P(Z,A)$. They are joined at $\emptyset$, a set that does not
contain $\Cc$, which Theorem~\ref{thm:join} excludes for minimal terms only. For a term that is not
minimal the test at $\Cc$ is therefore not conclusive, and here it would leave the summation in
$\sum_{A}P(A)\,P(Z\mid A,B)$, a correct expression, equal to $P(Z)$. For this reason the procedure
of Section~\ref{sec:algo} minimises every term that it writes after a summation.\footnote{An
example of this kind requires line~7 of ID. In the expressions that we examined, those returned
through line~6 alone do not have the depth of sums and ratios at which the effect appears, whereas
successive applications of lines~7 and~2 provide it. We will discuss this in a forthcoming paper.}

The procedure of TK removes the four summations as well, with a search in place of the minimisation. It sums $Y$
and $V$ as above, and for $X$ it joins $P(Z\mid A,B,X)$ with $P(X\mid B)$; here $\An(X)=\{A,B\}$ and
$\Jb^\pi\setminus\An^*(X)=\emptyset$, so the only candidate set is $\{A,B\}$, which is accepted,
and the summation leaves $P(Z\mid A,B)$, which TK does not minimise. For $A$ it joins
$P(Z\mid A,B)$ with $P(A)$; now $\An(A)=\emptyset$ and $\Jb^\pi\setminus\An^*(A)=\{B\}$, so the
candidate sets are $\emptyset$ and $\{B\}$. At $\emptyset$ the first test of TK is
$Z\indep B\mid A$, the statement that $B$ is idle, and the second is vacuous; the candidate is
accepted, the joint is $P(Z,A)$, and the result is $P(Z)$. The search thus performs at each join
the check that a minimisation performs once after each summation, and the single test of
Theorem~\ref{thm:join} replaces the search only if the terms are kept minimal.
 
In the second fraction no summation can be removed from the numerator, by either procedure. Its
denominator behaves like the expression above: with the minimisation it reduces to $P(Z)\,P(V)$,
without it the procedure stops at $\sum_{A}P(A)\,P(Z\mid A,B)\,P(V\mid A,B,Z)$, and TK returns
$P(Z)\,P(V\mid Z)$, in which $Z$ is idle.
\end{example}

\section{Insertion}
\label{sec:insert}

By Theorem~\ref{thm:join}, the join of the joint $P(\Jb\mid\Db)$ with the next term $P(V\mid\Cb)$
fails when one of the two independences \eqref{eq:testgraph} fails. When the second one fails,
$V$ depends, given $\Cb$, on a condition $M$ of the joint, $M\notin \Cb$
If $M$ lies above $V$, this dependence can be removed by turning $M$ into a head of the joint: a term
for $M$ is multiplied into the joint, and divided out after the summation, which requires it not to
depend on $S$. This is the insertion of \cite{tikkakarvanen2017}, and by Proposition~\ref{prop:reduce}
the only variables that need to be inserted are those of $\Nb$, the conditions of participating
terms that head none of them.

A participating term contains $S$, as its head or as a condition below its head, so every
participating head lies at or above $S$. The procedure inserts the variables of
$\Nb$ above $S$ and keeps the others as conditions:
\[
  \Rb:=\{x\in\Nb:\ x>S\},\qquad \Lb:=\{x\in\Nb:\ x<S\}.
\]
The final joint is then $P(\Hb\cup\Rb\mid\Lb)$, a reduced assembly in the sense of
Definition~\ref{def:fullreduced}. Let $h_1<\dots<h_m$ be its heads in topological order; the joint
is assembled from $h_m$ down to $h_1$, and the chain rule conditions each head on $\Lb$ and on the
heads below it:
\begin{equation}\label{eq:EM}
  \Eb_h:=\Lb\cup\{x\in\Hb\cup\Rb:\ x<h\},\qquad h\in\Hb\cup\Rb .
\end{equation}
Since $\Lb$ lies below $S$ and every head lies at or above $S$, this set consists of all the
variables of $\Hb\cup\Nb$ below $h$:
\[
  \Eb_h=\{x\in\Hb\cup\Nb:\ x<h\},\qquad h\in\Hb\cup\Rb .
\]

\begin{lemma}[Elimination through insertions]\label{lem:elim}
There exist terms $t_M$, with $M\in\Rb$, each a
conditional distribution of $M$ given variables of $\Eb_M$ and free of $S$, such that
\begin{equation}
\label{eq:decomp}
  \prod_{h\in\Hb}P(h\mid C_h)\prod_{M\in\Rb}t_M=P(\Hb\cup\Rb\mid\Lb)\;\;,  
\end{equation}
if and only if
\begin{enumerate}[label=(\alph*),nosep]
  \item $P(h\mid C_h)=P(h\mid\Eb_h)$ for every $h\in\Hb$, and
  \item $M\indep S\mid\Eb_M\setminus\{S\}$ for every $M\in\Rb$.
\end{enumerate}
In that case $t_M=P(M\mid\Eb_M)=P(M\mid\Eb_M\setminus\{S\})$ for every $M\in\Rb$, and
\begin{equation}\label{eq:elim}
  \sum_S\prod_{h\in\Hb}P(h\mid C_h)=\prod_{h\in\Hb\setminus\{S\}}P\bigl(h\mid\Eb_h\setminus\{S\}\bigr),
\end{equation}
where no term on the right-hand side contains $S$.
\end{lemma}

\begin{proof}
If the terms $t_M$ exist, every factor of the product is conditioned only on variables
below its head, so the highest head $h_m$
occurs only in its own factor.
Summing both sides over $h_m$ removes that factor by (R1) and shows that the product of the other
factors is $P(\{h_1,\dots,h_{m-1}\}\mid\Lb)$; dividing \eqref{eq:decomp} by this product (which positivity permits), the factor of
$h_m$ is $P(h_m\mid\Lb\cup\{h_1,\dots,h_{m-1}\})=P(h_m\mid\Eb_{h_m})$. Descending induction shows
that every factor is the one that the chain rule assigns to its head: for $h\in\Hb$ this is (a), and
for $M\in\Rb$ it gives $t_M=P(M\mid\Eb_M)$, which is free of $S$, that is, (b).

Conversely, under (a) and (b) the terms $t_M=P(M\mid\Eb_M)=P(M\mid\Eb_M\setminus\{S\})$ are free of
$S$, and by the chain rule and (a) they complete the participating terms to
$P(\Hb\cup\Rb\mid\Lb)$. Dividing by the terms $t_M$ and summing over $S$, which occurs neither in them nor in
$\Lb$, gives by (R2) and (R4)
\[
  \sum_S\prod_{h\in\Hb}P(h\mid C_h)
  =\frac{P\bigl((\Hb\cup\Rb)\setminus\{S\}\mid\Lb\bigr)}
        {\prod_{M\in\Rb}P(M\mid\Eb_M\setminus\{S\})}.
\]
The chain rule for the numerator assigns to each $h\neq S$ the set $\Eb_h\setminus\{S\}$, and the
factors with $h\in\Rb$ cancel against the denominator, which gives \eqref{eq:elim}.
\end{proof}

Condition~(a), on the participating terms, states that $h\indep\Eb_h\setminus C_h\mid C_h$, since
$C_h\subseteq\Eb_h$; in other words, a term must be independent on every variable of the joint below its head that its minimal conditioning set $C_h$ does not carry. Condition~(b), on the inserted terms, requires the term
inserted for $M$, which by the lemma is necessarily $P(M\mid\Eb_M)$, not to depend on $S$, so that
it can be divided out after the summation. 
As for the join set in Theorem~\ref{thm:join}, the conditioning set of an inserted term is fixed,
so that the insertion is decided at a single set, without a search: condition~(b) for the inserted
term, and condition~(a) for the terms already assembled, which must not depend on the conditions
that enter the joint with it.
The procedure of TK searches for this set among the sets $\An(M)\cup\mathbf{P}$,
$\mathbf{P}\subseteq\Jb^\pi\setminus\An^*(M)$; by the lemma, whatever set the search accepts, the
inserted term is $P(M\mid\Eb_M)$. The set $\Eb_M$ is determined by all the participating terms,
including those joined after $M$ (Example~\ref{ex:inset}). The lemma and its proof hold unchanged
for any $\Rb\subseteq\Nb$, with $\Lb=\Nb\setminus\Rb$ and the sets \eqref{eq:EM}.

The terms written by \eqref{eq:elim} are conditioned on $\Eb_h\setminus\{S\}$, below their heads, so the
procedure maintains the property assumed in Section~\ref{sec:setting}. They need not be minimal:
$\Eb_h\setminus\{S\}$ contains every variable of the joint below $h$, and the summation over $S$
can make some of them idle (Example~\ref{ex:restore}). Neither the lemma nor its proof uses
minimality; conditions (a) and (b) refer to the conditioning sets as they are written.\\
In the rest of the section we work out two examples. In
Example~\ref{ex:inset} the inserted term needs a condition that only a term joined later carries, which shows the importance of considering the full sets $\Rb$ and $\Lb$ \emph{ab initio}. Example~\ref{ex:frontdoor} shows again the importance of participation when insertions are used: a term of $\Oo_S$ cannot be joined, while its head can be inserted with another conditioning set.
 
\begin{example}[the conditioning set of an inserted term]\label{ex:inset}
Consider the graph below, with the topological order $v_0<v_1<\dots<v_5$, and the query
$P(v_2,v_5\mid \hat{v}_0,\hat{v}_4)$.
\begin{center}
\begin{tikzpicture}[node distance=11mm]
\node[vtx] (v0) {$v_0$};
\node[vtx,right=of v0] (v1) {$v_1$};
\node[vtx,right=of v1] (v2) {$v_2$};
\node[vtx,right=of v2] (v3) {$v_3$};
\node[vtx,right=of v3] (v4) {$v_4$};
\node[vtx,right=of v4] (v5) {$v_5$};
\draw[dir] (v0) -- (v1);
\draw[dir] (v1) -- (v2);
\draw[dir] (v2) -- (v3);
\draw[dir] (v3) -- (v4);
\draw[dir] (v4) -- (v5);
\draw[dir] (v1) to[bend left=35] (v5);
\draw[dir] (v2) to[bend left=30] (v4);
\draw[bid] (v0) to[bend right=30] (v3);
\end{tikzpicture}
\end{center}
ID returns
\[
  P(v_2,v_5\mid \hat{v}_0,\hat{v}_4)=\sum_{v_1}P(v_1\mid v_0)\,P(v_2\mid v_1)\,P(v_5\mid v_1,v_4).
\]
All three terms contain $v_1$, so that $\Hb=\{v_1,v_2,v_5\}$ and $\Nb=\{v_0,v_4\}$; hence
$\Rb=\{v_4\}$ and $\Lb=\{v_0\}$, and the variables are assembled in the order $v_5,v_4,v_2,v_1$.
\begin{enumerate}[label=\arabic*.,leftmargin=*,nosep]
  \item The term $P(v_5\mid v_1,v_4)$ forms the initial joint, with $\Jb=\{v_5\}$ and
    $\Db=\{v_1,v_4\}$.
  \item To insert the variable $v_4$, \eqref{eq:EM} gives $\Eb_{v_4}=\{v_0,v_1,v_2\}$, and
    condition~(b) requires $v_4\indep v_1\mid v_0,v_2$, which holds: the path $v_1\to v_2\to v_4$ is
    blocked at $v_2$, the path $v_1\leftarrow v_0\leftrightarrow v_3\to v_4$ at $v_0$, and the path
    $v_1\to v_5\leftarrow v_4$ at the collider $v_5$, which is not given. The inserted term is
    $P(v_4\mid v_0,v_2)$, which by (b) equals $P(v_4\mid v_0,v_1,v_2)$; it is joined at $\{v_0,v_1,v_2\}$,
    both independences holding, and the joint becomes $P(v_5,v_4\mid v_0,v_1,v_2)$.
  \item The join with $P(v_2\mid v_1)$ is fixed at $\Cc=\{v_0,v_1\}$; its second independence is
    $v_2\indep v_0\mid v_1$, which holds, and the joint becomes $P(v_5,v_4,v_2\mid v_0,v_1)$.
  \item The join with $P(v_1\mid v_0)$ at $\Cc=\{v_0\}$ involves only vacuous independences, and
    the joint becomes $P(v_5,v_4,v_2,v_1\mid v_0)$.
  \item Summing over $v_1$ and factorising in the order of assembly gives
    $P(v_5\mid v_0,v_2,v_4)\,P(v_4\mid v_0,v_2)\,P(v_2\mid v_0)$; the factor of $v_4$ is the
    inserted term and is divided out, so that
    \[
      P(v_2,v_5\mid \hat{v}_0,\hat{v}_4)=P(v_2\mid v_0)\,P(v_5\mid v_0,v_2,v_4).
    \]
\end{enumerate}
Without the insertion, the join of $P(v_5\mid v_1,v_4)$ with $P(v_2\mid v_1)$ would fail, since
$v_2\nindep v_4\mid v_1$ because of $v_2\to v_4$: the insertion of $v_4$ is forced. The condition $v_0$ of the
inserted term comes from $P(v_1\mid v_0)$, the last term to be joined, which neither the joint
$P(v_5\mid v_1,v_4)$ nor $P(v_2\mid v_1)$ mentions. Had the term for $v_4$ been
conditioned on $\Db\setminus\{v_4\}=\{v_1\}$, condition~(b) would have required $v_4\indep v_1$; had
it been conditioned on the set $\Cc$ of the failed join together with $v_2$, it would have
required $v_4\indep v_1\mid v_2$. Both fail, because the path
$v_4\leftarrow v_3\leftrightarrow v_0\to v_1$ is open, and $v_4$ could not have been inserted.
 
The procedure of TK reaches the same expression: after its two candidate sets
for the first join have failed, it tries the missing variable $v_4$ with the sets
$\An(v_4)\cup\mathbf{P}$, the first of which, $\An(v_4)=\{v_0,v_1,v_2\}$, coincides with $\Eb_{v_4}$. In
general however the sets $\Eb$ are not defined by ancestry, and \eqref{eq:EM} gives them without a search.
\end{example}
 
\begin{example}[a retained term between the participating terms]\label{ex:frontdoor}
Consider the graph below, with the topological order $v_0<v_1<v_2<v_3<v_4$, and the query
$P(v_2,v_4\mid \hat{v}_0)$.
\begin{center}
\begin{tikzpicture}[node distance=12mm]
\node[vtx] (v0) {$v_0$};
\node[vtx,right=of v0] (v1) {$v_1$};
\node[vtx,right=of v1] (v2) {$v_2$};
\node[vtx,right=of v2] (v3) {$v_3$};
\node[vtx,right=of v3] (v4) {$v_4$};
\draw[dir] (v1) -- (v2);
\draw[dir] (v2) -- (v3);
\draw[dir] (v3) -- (v4);
\draw[dir] (v0) to[bend left=40] (v4);
\draw[dir] (v1) to[bend left=30] (v4);
\draw[bid] (v0) to[bend right=30] (v3);
\end{tikzpicture}
\end{center}
ID returns
\[
  P(v_2,v_4\mid \hat{v}_0)=\sum_{v_1,v_3}P(v_1)\,P(v_2\mid v_1)\,P(v_3\mid v_2)\,P(v_4\mid v_0,v_1,v_3).
\]
The summation variables are tried in decreasing order. The participating terms of $v_3$ are
$P(v_4\mid v_0,v_1,v_3)$ and $P(v_3\mid v_2)$; their join is fixed at $\{v_0,v_1,v_2\}$, and its
second independence, $v_3\indep\{v_0,v_1\}\mid v_2$, fails because of $v_0\leftrightarrow v_3$. No
variable of $\Nb=\{v_0,v_1,v_2\}$ lies above $v_3$, so nothing is inserted, and $v_3$ cannot be
eliminated: $v_0$ lies below $v_3$, and no insertion removes it from $\Eb_{v_3}$.\\
The participating terms of $v_1$ are $P(v_4\mid v_0,v_1,v_3)$, $P(v_2\mid v_1)$ and
$P(v_1)$, while $P(v_3\mid v_2)$, the term that could not be joined, remains outside the
summation; here $\Hb=\{v_1,v_2,v_4\}$ and $\Nb=\{v_0,v_3\}$, so that $\Rb=\{v_3\}$ and
$\Lb=\{v_0\}$, and the order of assembly is $v_4,v_3,v_2,v_1$.
\begin{enumerate}[label=\arabic*.,leftmargin=*,nosep]
  \item The term $P(v_4\mid v_0,v_1,v_3)$ forms the initial joint, with $\Jb=\{v_4\}$ and
    $\Db=\{v_0,v_1,v_3\}$.
  \item To insert the variable $v_3$, \eqref{eq:EM} gives $\Eb_{v_3}=\{v_0,v_1,v_2\}$, and
    condition~(b) requires $v_3\indep v_1\mid v_0,v_2$, which holds: the path $v_1\to v_2\to v_3$ is
    blocked at $v_2$, and the paths through $v_4$ are blocked at the collider $v_4$, which is not
    given. The inserted term is $P(v_3\mid v_0,v_2)$, which by (b) equals $P(v_3\mid v_0,v_1,v_2)$; it is joined at
    $\{v_0,v_1,v_2\}$, both independences holding, and the joint becomes
    $P(v_4,v_3\mid v_0,v_1,v_2)$.
  \item The join with $P(v_2\mid v_1)$ is fixed at $\Cc=\{v_0,v_1\}$; its second independence,
    $v_2\indep v_0\mid v_1$, holds, and the joint becomes $P(v_4,v_3,v_2\mid v_0,v_1)$.
  \item The join with $P(v_1)$ at $\Cc=\{v_0\}$ requires $v_1\indep v_0$, which holds, and the
    joint becomes $P(v_4,v_3,v_2,v_1\mid v_0)$.
  \item Summing over $v_1$ and factorising in the order of assembly gives
    $P(v_4\mid v_0,v_2,v_3)\,P(v_3\mid v_0,v_2)\,P(v_2)$, where $P(v_2\mid v_0)$ has been
    minimised to $P(v_2)$; the factor of $v_3$ is the inserted term and is divided out. With
    $P(v_3\mid v_2)$ restored by (R2),
    \[
      P(v_2,v_4\mid \hat{v}_0)=\sum_{v_3}P(v_2)\,P(v_3\mid v_2)\,P(v_4\mid v_0,v_2,v_3).
    \]
\end{enumerate}
Here the insertion of $v_3$ is forced: without it, the join of $P(v_4\mid v_0,v_1,v_3)$ with
$P(v_2\mid v_1)$ would fail, since $v_2\nindep v_3\mid v_1$ because of $v_2\to v_3$.
A further attempt on $v_3$ fails as before, since the join of $P(v_4\mid v_0,v_2,v_3)$ with
$P(v_3\mid v_2)$ at $\{v_0,v_2\}$ requires $v_3\indep v_0\mid v_2$. The term $P(v_3\mid v_2)$
cannot be joined because of $v_0\leftrightarrow v_3$; left outside the summation, it leaves its
head free to be inserted with the conditioning set $\{v_0,v_1,v_2\}$ that the chain rule
requires, and the inserted term $P(v_3\mid v_0,v_2)$ differs from it by the condition
$v_0$.
 
For TK the term $P(v_3\mid v_2)$ takes part in the elimination of $v_3$ and, as a retained term,
in that of $v_1$. To eliminate $v_3$, it joins
$P(v_4\mid v_0,v_1,v_3)$ with $P(v_3\mid v_2)$, and both candidate sets, $\{v_1,v_2\}$ and
$\{v_0,v_1,v_2\}$, fail; no variable is missing. To eliminate $v_1$, it must assemble $v_4$,
$v_3$, $v_2$ and $v_1$, beginning with the same join, which fails again, and no variable between
$v_1$ and $v_4$ is missing, since $v_2$ and $v_3$ head terms and $v_0$ lies below $v_1$, so that the
expression is left unchanged. In the topological order $v_1<v_2<v_3<v_0<v_4$, however, $v_0$ lies
between $v_1$ and $v_4$ and heads no term; the procedure then inserts it and eliminates $v_1$ as
well.
\end{example}

\section{The procedure}
\label{sec:algo}

The previous sections determine every step of an elimination: the terms to assemble
(Proposition~\ref{prop:reduce}), the set of each join (Theorem~\ref{thm:join}), and the term of each
insertion (Lemma~\ref{lem:elim}). This section combines them into a procedure that keeps the
structure of TK, a loop over the summation variables that calls a join, which calls an insertion
where one is required, followed by a factorisation, so that the two procedures can be compared step
by step; the difference is that every set is computed and none is searched for.

Algorithms~\ref{alg:simplify}--\ref{alg:minimize} state the procedure, under the names used in
\cite{tikkakarvanen2017}. A term $P(V\mid\Cb)$ is written $\langle V,\Cb\rangle$, and a joint
$P(\Jb\mid\Db)$ is written $\langle\Jb,\Db\rangle$, with conditioning set $\Db$ and heads $\Jb$
kept as a sequence in the order of assembly; $(\Jb,V)$ denotes the sequence $\Jb$ followed by
$V$, and in set operations and independence statements $\Jb$ stands for the set of its elements.
$\Fail$ marks an unsuccessful call, and $(\cdot\indep\cdot\mid\cdot)_G$ a d-separation test in $G$
with its latent variables, true when either of its first two arguments is empty or contained in the
third, and false when they share a variable outside the third.

\begin{algorithm}[htbp]
\caption{Simplification of an atomic expression $\sum_{\Ss}\prod_{t\in\Tb}t$ given graph $G$ and
topological ordering $\pi$.}\label{alg:simplify}
\begin{algorithmic}[1]
\Function{Simplify}{$\Tb,\Ss,G,\pi$}
  \State $j\gets0$
  \While{$j<|\Ss|$}
    \State $j\gets j+1$ \Comment{$S_j$: the $j$-th of $\Ss$, in decreasing order of $\pi$}
    \State $\Pp\gets\{\langle V,\Cb\rangle\in\Tb: V=S_j\text{ or }S_j\in\Cb\}$;\ \
           $\Oo\gets\Tb\setminus\Pp$ \Comment{participating terms}
    \State $\Hb\gets\{V:\langle V,\Cb\rangle\in\Pp\}$;\ \
           $\Nb\gets\bigl(\bigcup_{\langle V,\Cb\rangle\in\Pp}\Cb\bigr)\setminus\Hb$
    \State $\Rb\gets\{x\in\Nb: x>S_j\}$;\ \ $\Lb\gets\Nb\setminus\Rb$ \Comment{inserted variables}
    \State $\Jb\gets()$;\ \ $\Db\gets\emptyset$;\ \ $\mathbf{Q}\gets\emptyset$;\ \ $k\gets1$
    \While{$k\le|\Pp|$} \Comment{$\langle V_k,\Cb_k\rangle$: the $k$-th term of $\Pp$, by decreasing head}
      \State $r\gets\Call{Join}{\Jb,\Db,V_k,\Cb_k,\Hb,\Rb,\Lb,S_j,G}$
      \If{$r=\Fail$} \textbf{break} \EndIf
      \State $\langle\Jb,\Db,\mathbf{Q}'\rangle\gets r$;\ \ $\mathbf{Q}\gets\mathbf{Q}\cup\mathbf{Q}'$;\ \
             $k\gets k+1$
    \EndWhile
    \If{$k>|\Pp|$} \Comment{$\prod\Pp\cdot\prod\mathbf{Q}=P(\Jb\mid\Db)$}
      \State $\Tb\gets\Oo\cup\Call{Factorize}{\Jb,\Db,S_j,\mathbf{Q},G}$
        \Comment{sum out $S_j$, divide by $\mathbf{Q}$}
      \State $\Ss\gets\Ss\setminus\{S_j\}$;\ \ $j\gets0$ \Comment{restart on the new terms}
    \EndIf
  \EndWhile
  \State \Return $\langle\Tb,\Ss\rangle$
\EndFunction
\end{algorithmic}
\end{algorithm}

For each summation variable $S_j$, lines~5--7 determine the participating terms, their heads
$\Hb$, the set $\Nb$ of Definition~\ref{def:part}, and the inserted variables $\Rb$ of
Section~\ref{sec:insert}; the terms of $\Oo$ are put back unchanged after the summation, by (R2).
The loop of lines~9--12 takes the participating terms in decreasing order of their heads and calls
\Proc{Join} for each; \Proc{Join} calls \Proc{Insert} for the variables of $\Rb$ that lie above the
head, so that the participating heads and the inserted variables are assembled in decreasing order.
When the loop completes, the participating terms times the inserted terms equal $P(\Jb\mid\Db)$, and
\Proc{Factorize} sums out $S_j$ and divides by the inserted terms (line~14); $S_j$ is removed, and
the summation variables are tried again from the first, since one elimination may enable another.

\begin{algorithm}[htbp]
\caption{Join of $P(\Jb\mid\Db)$ with $P(V\mid\Cb)$ at the set $\Cc$ of Theorem~\ref{thm:join}, after
the insertion of the variables of $\Rb$ above $V$.}
\label{alg:join}
\begin{algorithmic}[1]
\Function{Join}{$\Jb,\Db,V,\Cb,\Hb,\Rb,\Lb,S,G$}
  \If{$\Jb=()$} \Return $\langle(V),\Cb,\emptyset\rangle$ \Comment{the initial joint}
  \EndIf
  \State $\mathbf{Q}'\gets\emptyset$
  \For{$M\in\Db$ with $M>V$, in decreasing order} \Comment{conditions above $V$: inserted first}
    \State $r\gets\Call{Insert}{\Jb,\Db,M,\Hb,\Rb,\Lb,S,G}$
    \If{$r=\Fail$} \Return $\Fail$ \EndIf
    \State $\langle\Jb,\Db,q\rangle\gets r$;\ \ $\mathbf{Q}'\gets\mathbf{Q}'\cup\{q\}$
  \EndFor
  \State $\Cc\gets(\Db\cup\Cb)\setminus\{V\}$ \Comment{the set of Theorem~\ref{thm:join}}
  \If{$\bigl(\Jb\indep(\Cb\cup\{V\})\setminus\Db\mid\Db\bigr)_G$
      \textbf{and} $\bigl(V\indep(\Db\setminus\{V\})\setminus\Cb\mid\Cb\bigr)_G$}
    \State \Return $\langle(\Jb,V),\Cc,\mathbf{Q}'\rangle$ \Comment{$P(\Jb,V\mid\Cc)$}
  \EndIf
  \State \Return $\Fail$
\EndFunction
\end{algorithmic}
\end{algorithm}

\Proc{Join} returns the term itself when the joint is empty (line~2). Otherwise it first inserts
the conditions of the joint that lie above $V$, from the highest down (lines~4--7). These are the
variables of $\Rb$ above $V$, and each must become a head before $V$ is joined, so that every head
is conditioned only on variables below it. It then tests the two independences of
Theorem~\ref{thm:join} at the set $\Cc$ and nowhere else (lines~8--9). By Theorem~\ref{thm:join}(ii) a
join of minimal terms produces a minimal joint.

Lines~4--7 insert every condition $M$ of the joint that lies above $V$, whether or not the
insertion is necessary. It is necessary when a term $P(V\mid\Cb)$ with head below $M$ depends on
$M$ without having it among its conditions ($V\nindep M\mid\Cb$, so that the second independence
of \eqref{eq:testgraph} fails), and otherwise each remaining join would succeed with $M$ kept as a
condition ($V\indep M\mid\Cb$, as $a\indep b$ in \eqref{eq:ex-join}), which is how TK proceeds,
inserting only after a join has failed. We insert in both cases because this fixes the conditioning
set of each inserted term in advance, at no cost in eliminations (Corollary~\ref{cor:complete}).
\begin{algorithm}[htbp]
\caption{Insertion of a term for $M$ into the joint $P(\Jb\mid\Db)$.}
\label{alg:insert}
\begin{algorithmic}[1]
\Function{Insert}{$\Jb,\Db,M,\Hb,\Rb,\Lb,S,G$}
  \State $\Eb\gets\bigl(\Lb\cup\{x\in\Hb\cup\Rb: x<M\}\bigr)\setminus\{S\}$
    \Comment{the set \eqref{eq:EM}, without $S$}
  \If{\textbf{not} $\bigl(M\indep S\mid\Eb\bigr)_G$} \Return $\Fail$ \Comment{condition (b)}
  \EndIf
  \If{\textbf{not} $\bigl(\Jb\indep\Eb\setminus\Db\mid\Db\bigr)_G$} \Return $\Fail$
    \Comment{condition (a) for the heads of $\Jb$}
  \EndIf
  \State \Return $\bigl\langle(\Jb,M),\ (\Db\cup\Eb)\setminus\{M\},\ \langle M,\Eb\rangle\bigr\rangle$
\EndFunction
\end{algorithmic}
\end{algorithm}

\Proc{Insert} computes the conditioning set \eqref{eq:EM} without $S$, fixed by
Lemma~\ref{lem:elim} (line~2), and performs two tests. The first is condition~(b): the inserted
term $P(M\mid\Eb)$ does not depend on $S$ (line~3). The second asks that the joint assembled so far
keep its value when it is conditioned on $\Eb$ as well (line~4): the terms already absorbed may not
depend on the variables below $M$ that enter the joint with the inserted term, which is
condition~(a) for their heads. When $M$ is inserted, $M$ and $S$ are conditions of the joint and
every other condition lies in $\Eb$, so that the new conditioning set
$(\Db\cup\Eb)\setminus\{M\}$ is $\Eb\cup\{S\}$ (line~5). The two tests are those of
Theorem~\ref{thm:join} for the join of $P(\Jb\mid\Db)$ with $P(M\mid\Eb)$ at this set, and those of
the insertion of TK at the single candidate $\Eb\cup\{S\}$. The inserted term is not minimised:
its conditioning set is fixed by the chain rule, and the term is later divided out.

\begin{algorithm}[htbp]
\caption{Summation over $S$ of the joint $P(\Jb\mid\Db)$, inserted terms divided out.}
\label{alg:factorize}
\begin{algorithmic}[1]
\Function{Factorize}{$\Jb,\Db,S,\mathbf{Q},G$}
  \State $\Tb'\gets\emptyset$ \Comment{$\Jb=(j_m,\dots,j_1)$ in the order of assembly, $j_m>\dots>j_1$}
  \For{$i=m,\dots,1$}
    \If{$j_i=S$ \textbf{or} a term with head $j_i$ is in $\mathbf{Q}$}
      \State \textbf{continue} \Comment{summed out, or divided out}
    \EndIf
    \State $\Tb'\gets\Tb'\cup\bigl\{\bigl\langle j_i,\ \Call{Minimize}{\{j_i\},\
           (\Db\cup\{j_1,\dots,j_{i-1}\})\setminus\{S\},\ G}\bigr\rangle\bigr\}$
      \Comment{its factor, minimal}
  \EndFor
  \State \Return $\Tb'$
\EndFunction
\end{algorithmic}
\end{algorithm}

\Proc{Factorize} carries out the summation by the argument of Lemma~\ref{lem:elim}. The joint
assembled by the loop is a product of the participating terms and of the inserted terms, and each
of them is the factor that the chain rule assigns to its head, that is, the factor of $j_i$ given
$\Db$ and $j_{i+1},\dots,j_m$. Summing over $S$ leaves a joint distribution of the other heads,
whose chain rule assigns to $j_i$ the set $(\Db\cup\{j_{i+1},\dots,j_m\})\setminus\{S\}$. For an
inserted head this factor equals the inserted term, which does not depend on $S$; it cancels
against the division by that term, and the head is skipped (lines~4--5). Every other factor is
returned with a minimal conditioning set (line~6).

\begin{algorithm}[htbp]
\caption{A minimal conditioning set for a term or joint with heads $\Eh$.}\label{alg:minimize}
\begin{algorithmic}[1]
\Function{Minimize}{$\Eh,\Cb,G$}
  \Repeat
    \State $removed\gets\Proc{false}$
    \For{$c\in\Cb$}
      \If{$\bigl(\Eh\indep c\mid\Cb\setminus\{c\}\bigr)_G$}
        \Comment{$c$ is idle: $P(\Eh\mid\Cb)=P(\Eh\mid\Cb\setminus\{c\})$}
        \State $\Cb\gets\Cb\setminus\{c\}$;\ \ $removed\gets\Proc{true}$;\ \ \textbf{break}
      \EndIf
    \EndFor
  \Until{$removed=\Proc{false}$}
  \State \Return $\Cb$ \Comment{no single condition can be dropped}
\EndFunction
\end{algorithmic}
\end{algorithm}

\Proc{Minimize} drops idle conditions one at a time; dropping one may make another droppable, so
it repeats until none can be dropped, and the result is minimal in the sense of
Definition~\ref{def:min}. Only \Proc{Factorize} calls it, for every term it returns: a condition
left idle by the summation would be read as a condition at the next join, and the test at $\Cc$
could then fail for terms that can be joined (Example~\ref{ex:restore}).

The next proposition states that the procedure is correct, and that it decides the two conditions
of Lemma~\ref{lem:elim}: the loop of \Proc{Simplify} completes whenever the summation can be
carried out through the joint $P(\Hb\cup\Rb\mid\Lb)$, and the expression that it then writes
equals the original one.

\begin{proposition}\label{prop:algo}
If the participating terms are minimal and the loop of lines~9--12 of \Proc{Simplify} completes
for $S_j$, line~14 replaces the terms by an expression equal to $\sum_{S_j}\prod_{t\in\Tb}t$, in
which $S_j$ does not occur and every term is minimal. Conversely, if conditions (a) and (b) of
Lemma~\ref{lem:elim} hold for the set $\Rb$ of line~7, the loop completes.
\end{proposition}

\begin{proof}
Each absorbed term is the factor that the chain rule, in the order of assembly, assigns to its
head. Indeed, suppose this holds for the joint $P(\Jb\mid\Db)$, and let \Proc{Join} absorb the next
term at $\Cc$, so that $\Jb\indep\mathbf{W}\mid\Db$ with $\mathbf{W}=(\Cb\cup\{V\})\setminus\Db$.
For a head $a$ of $\Jb$ and the set $\mathbf{Y}$ of heads absorbed after it, decomposition gives
$\{a\}\cup\mathbf{Y}\indep\mathbf{W}\mid\Db$, and symmetry and weak union give
$a\indep\mathbf{W}\mid\Db\cup\mathbf{Y}$, so that the factor of $a$ given
$\Db\cup\mathbf{Y}$ is also its factor given $\Cc\cup\{V\}\cup\mathbf{Y}$, which is the set that the
chain rule assigns to it after the join; the new term is the factor of $V$ given $\Cc$ by the
second independence. An insertion is the same step for the term $P(M\mid\Eb)$. When \Proc{Insert}
is called for $M$, the heads of the joint are the variables of $\Hb\cup\Rb$ above $M$; $M$ and
$S_j$ are conditions of participating terms already absorbed, and every other condition of the
joint lies below $M$, hence in $\Eb$. The term is thus absorbed at
$(\Db\cup\Eb)\setminus\{M\}=\Eb\cup\{S_j\}$; line~4 of \Proc{Insert} is the first independence,
with $\mathbf{W}=\Eb\setminus\Db$, and line~3 is the second, since
$(\Db\setminus\{M\})\setminus\Eb=\{S_j\}$. When the loop completes, the participating terms times
the inserted terms therefore equal $P(\Jb\mid\Db)$; the inserted terms are free of $S_j$, so they
leave the summation by (R2), and the summation removes $S_j$ from the heads. \Proc{Factorize}
writes the result by the chain rule, divides by the inserted terms as explained above, and
minimises; \Proc{Minimize} drops only idle conditions, and the terms of $\Oo$ are restored by (R2).

Conversely, under (a) and (b) the factors $P(h\mid\Eb_h)$ of Lemma~\ref{lem:elim}, taken from the
highest variable of $\Hb\cup\Rb$ down to any variable $g$, multiply to the joint distribution of
these variables given $\Eb_g$, so that the joint assembled when $g$ has been absorbed satisfies
$P(\Jb\mid\Db)=P(\Jb\mid\Eb_g)$, with $\Db\subseteq\Eb_g$. The next variable to be absorbed, $V$,
is the next one below $g$ in $\Hb\cup\Rb$, so that $\Eb_g=\Eb_V\cup\{V\}$. The first test of
\Proc{Join}, and line~4 of \Proc{Insert}, ask that $\Jb$ be independent, given $\Db$, of variables
of $\Eb_g\setminus\Db$, which follows by decomposition. The second test of \Proc{Join} asks that
$V$ be independent, given $\Cb$, of variables of $\Eb_V\setminus\Cb$, which follows from (a) by
decomposition, and line~3 of \Proc{Insert} is (b); hence every test succeeds, and the loop
completes.
\end{proof}

The procedure makes no search: every set it tests is computed, so its cost is polynomial, whereas a
join of TK may examine up to $2^{|\Jb^\pi\setminus\An^*(V)|}$ candidate sets
(Example~\ref{ex:cost}). We count
d-separation tests, each of which takes time linear in the size of $G$ \cite{geigervermapearl1990}.

\begin{proposition}[Cost]\label{prop:cost}
Let $G$ have $n$ variables, and let one attempt, for one summation variable, assemble $m=|\Hb\cup\Rb|$
heads. The attempt performs at most $2m$ tests in \Proc{Join} and \Proc{Insert} and at most $m-1$
calls of \Proc{Minimize}, each of which performs at most $c(c+1)$ tests for a set of $c\le n$
conditions. A run with $k$ summation variables makes at most $k(k+1)/2$ attempts, and therefore
$O(k^2n^3)$ tests in all.
\end{proposition}

\begin{proof}
The loop calls \Proc{Join}, with two tests, once for each participating term, and \Proc{Join} calls
\Proc{Insert}, with two tests, once for each variable of $\Rb$, so that there are at most
$2(|\Hb|+|\Rb|)=2m$ tests. \Proc{Factorize} calls \Proc{Minimize} once for each head of the joint
that is neither summed out nor divided out, that is, for each participating head other than $S_j$;
since $S_j$ is itself one of the $m$ heads, there are at most $|\Hb|-1\le m-1$ calls. Each pass of the loop of \Proc{Minimize} performs at most
$c$ tests and either removes a condition or ends the call, so there are at most $c+1$ passes.
Between two successful eliminations, \Proc{Simplify} tries each remaining summation variable at
most once, so a run makes at most $k+(k-1)+\dots+1$ attempts; since $m\le n$, each attempt makes
$O(n^3)$ tests.
\end{proof}

The next example traces a whole run through the algorithms, on a graph for which TK removes no
summation in any topological order.

\begin{example}[a run of the procedure]\label{ex:robust}
Consider the graph below, with the topological order $v_0<v_1<\dots<v_5$, and the query
$P(v_4,v_5\mid \hat{v}_2)$.
\begin{center}
\begin{tikzpicture}[node distance=12mm]
\node[vtx] (v0) {$v_0$};
\node[vtx,right=of v0] (v1) {$v_1$};
\node[vtx,right=of v1] (v2) {$v_2$};
\node[vtx,right=of v2] (v3) {$v_3$};
\node[vtx,right=of v3] (v4) {$v_4$};
\node[vtx,right=of v4] (v5) {$v_5$};
\draw[dir] (v0) -- (v1);
\draw[dir] (v1) -- (v2);
\draw[dir] (v1) to[bend left=30] (v3);
\draw[dir] (v1) to[bend left=38] (v4);
\draw[dir] (v1) to[bend left=46] (v5);
\draw[dir] (v2) to[bend left=30] (v5);
\draw[dir] (v3) to[bend left=22] (v5);
\draw[bid] (v0) to[bend right=30] (v2);
\draw[bid] (v0) to[bend right=40] (v4);
\draw[bid] (v3) -- (v4);
\end{tikzpicture}
\end{center}
ID returns
\[
  P(v_4,v_5\mid \hat{v}_2)=\sum_{v_0,v_1,v_3}P(v_0)\,P(v_1\mid v_0)\,P(v_3\mid v_1)\,
  P(v_4\mid v_0,v_1,v_3)\,P(v_5\mid v_1,v_2,v_3),
\]
and \Proc{Simplify} tries the summation variables in the order $v_3,v_1,v_0$.

\emph{$v_3$.} Line~5 selects the terms that contain $v_3$, namely $P(v_3\mid v_1)$,
$P(v_4\mid v_0,v_1,v_3)$ and $P(v_5\mid v_1,v_2,v_3)$. Lines~6--7 give $\Hb=\{v_3,v_4,v_5\}$ and
$\Nb=\{v_0,v_1,v_2\}$, which lies below $v_3$, so that $\Rb=\emptyset$ and $\Lb=\Nb$. The loop begins
with $v_5$, whose term is the initial joint $\langle(v_5),\{v_1,v_2,v_3\}\rangle$, and calls
\Proc{Join} for $P(v_4\mid v_0,v_1,v_3)$ at $\Cc=\{v_0,v_1,v_2,v_3\}$. The first test,
$v_5\indep\{v_0,v_4\}\mid v_1,v_2,v_3$, holds; the second, $v_4\indep v_2\mid v_0,v_1,v_3$, fails,
because the path $v_4\leftrightarrow v_0\leftrightarrow v_2$ is open when its collider $v_0$ is
given, and the loop breaks at line~11.

\emph{$v_1$.} Now $P(v_1\mid v_0)$ participates as well, $\Hb=\{v_1,v_3,v_4,v_5\}$,
$\Nb=\{v_0,v_2\}$, $\Rb=\{v_2\}$ and $\Lb=\{v_0\}$. The loop again begins with $v_5$ and $v_4$, and
the same join fails: $v_2$, which lies below $v_4$, is inserted only later, and is still a condition
of the joint when $v_4$ is joined.

\emph{$v_0$.} Line~5 selects $P(v_0)$, $P(v_1\mid v_0)$ and $P(v_4\mid v_0,v_1,v_3)$, and puts
$P(v_3\mid v_1)$ and $P(v_5\mid v_1,v_2,v_3)$ in $\Oo$. Then $\Hb=\{v_0,v_1,v_4\}$,
$\Nb=\{v_3\}$, $\Rb=\{v_3\}$ and $\Lb=\emptyset$, and the loop calls \Proc{Join} for the terms of
$v_4$, $v_1$ and $v_0$.
\begin{enumerate}[label=\arabic*.,leftmargin=*,nosep]
  \item The term for $v_4$ is the initial joint $\langle(v_4),\{v_0,v_1,v_3\}\rangle$.
  \item Called for $P(v_1\mid v_0)$, \Proc{Join} first inserts $v_3$, the variable of $\Rb$ above
    $v_1$. \Proc{Insert} computes $\Eb=\{v_1\}$ (line~2) and tests $v_3\indep v_0\mid v_1$
    (line~3), which holds: $v_3\leftarrow v_1\leftarrow v_0$ is blocked at $v_1$, and
    $v_3\leftrightarrow v_4\leftrightarrow v_0$ at the collider $v_4$, which is not given. The test
    of line~4 is vacuous, since $\Eb\subseteq\Db$. The joint is
    $\langle(v_4,v_3),\{v_0,v_1\}\rangle$, and the inserted term is $\langle v_3,\{v_1\}\rangle$.
  \item \Proc{Join} then adds $P(v_1\mid v_0)$ at $\{v_0\}$ and, called for the last term, $P(v_0)$
    at $\emptyset$, with vacuous tests only; the joint is
    $\langle(v_4,v_3,v_1,v_0),\emptyset\rangle$.
  \item \Proc{Factorize} (line~14) writes $v_4$ with \Proc{Minimize}$(\{v_4\},\{v_1,v_3\})=\{v_1,v_3\}$,
    skips $v_3$, which is in $\mathbf{Q}$, writes $v_1$ with the empty set, and skips $v_0$. With the
    terms of $\Oo$ restored,
    \[
      P(v_4,v_5\mid \hat{v}_2)=\sum_{v_1,v_3}P(v_1)\,P(v_3\mid v_1)\,P(v_4\mid v_1,v_3)\,
      P(v_5\mid v_1,v_2,v_3).
    \]
\end{enumerate}
After line~15 has removed $v_0$, the new attempts on $v_3$ and $v_1$ again begin with the join of
$P(v_5\mid v_1,v_2,v_3)$ and $P(v_4\mid v_1,v_3)$, now at $\{v_1,v_2,v_3\}$, whose second test
$v_4\indep v_2\mid v_1,v_3$ fails, since the collider $v_0$ on $v_4\leftrightarrow v_0\leftrightarrow
v_2$ has the given descendant $v_1$; the run ends with the expression above. The insertion of $v_3$ is
forced: without it, the join of $P(v_1\mid v_0)$ in step~3 would have failed, since
$v_1\nindep v_3\mid v_0$ because of $v_1\to v_3$.

TK must assemble $v_5$, $v_4$ and $v_3$ to eliminate
$v_3$, and its join of $P(v_5\mid v_1,v_2,v_3)$ with $P(v_4\mid v_0,v_1,v_3)$ rejects all four
candidate sets $\{v_0,v_1\}\cup\mathbf{P}$, $\mathbf{P}\subseteq\{v_2,v_3\}$; no variable is
missing. To eliminate $v_1$ or $v_0$, it must assemble the same two terms first, the same join
fails, and the missing variable $v_2$ cannot be inserted with any of its candidate sets. The
expression is left unchanged, in each of the eight topological orders of the graph, whereas the
elimination of $v_0$ above succeeds in all of them.
\end{example}

\Proc{Simplify} tries the summation variables in a fixed order, from the highest down, and starts
again from the top after each elimination. The result of a run can depend on this order, since an
elimination can prevent a later one (Example~\ref{ex:order}). A variant of the procedure could
explore several orders, for instance by backtracking over the eliminations that succeed and keeping the
expression with the fewest summations. Each attempt keeps the cost of Proposition~\ref{prop:cost}, but
the number of orders can grow as $k!$ (in Example~\ref{ex:order} such an exploration recovers the
expression that TK reaches). For simplicity, We
have not included such a search in the algorithms above.

\section{Soundness and completeness}
\label{sec:sc}

This section collects what the previous ones establish about the procedure as a whole. It never
writes a wrong expression, and for a single summation variable it does everything that joins and
insertions can do: it eliminates the variable whenever some assembly eliminates it. We then state
what remains open, the existence of steps of another kind and the order in which the summation
variables are tried.

The procedure is sound: it rewrites the expression through identities only. Terms free of $S$ leave
the summation by (R2); a join is the chain rule, valid exactly when the independences
\eqref{eq:testgraph} hold (Theorem~\ref{thm:join}); an inserted term is positive and free of $S$, so
that multiplying by it and dividing by it after the summation changes nothing; the summation is a
marginalisation by (R1) and (R4); and minimisation drops only idle conditions.

\begin{corollary}\label{cor:complete}
If some assembly eliminates $S$ from an atomic expression of the ID algorithm, the procedure of this paper
eliminates $S$ from that expression.
\end{corollary}

\begin{proof}
As noted after Proposition~\ref{prop:reduce}, the assembly may be taken in decreasing
topological order. By Proposition~\ref{prop:reduce}, its reduction eliminates $S$ as well: the
participating terms are assembled with inserted terms for a set $\Rb'\subseteq\Nb$, the variables
of $\Lb'=\Nb\setminus\Rb'$ being conditions. By Lemma~\ref{lem:elim}, which holds for any such
choice, conditions (a) and (b) hold for $\Rb'$, with the sets
$\Eb'_h=\Lb'\cup\{x\in\Hb\cup\Rb':x<h\}$. To pass from $\Rb'$ to the set $\Rb$ of the procedure, observe for
(a) that the set $\Eb_h$ consists of the variables of $\Hb\cup\Nb$ below $h$, all of which belong to
$\Eb'_h$, so that $P(h\mid C_h)=P(h\mid\Eb'_h)$ gives $P(h\mid C_h)=P(h\mid\Eb_h)$ by decomposition.
For (b), the chain rule and (a) give
\[
  \prod_{M\in\Rb}P(M\mid\Eb_M)=\frac{P(\Hb\cup\Rb\mid\Lb)}{\prod_{h\in\Hb}P(h\mid C_h)}
  =\frac{P(\Hb\cup\Nb)}{P(\Lb)\prod_{h\in\Hb}P(h\mid C_h)},
\]
and the same identity holds for $\Rb'$ and $\Lb'$. The two products of inserted terms thus differ by
the factor $P(\Lb')/P(\Lb)$, which does not contain $S$, and the product for $\Rb'$ is free of $S$ by (b).
Hence the product for $\Rb$ is free of $S$, and so is each of its factors, by the argument that
ends the proof of Proposition~\ref{prop:reduce} in Appendix~\ref{app:step1}; this is (b). By
Lemma~\ref{lem:elim} and Proposition~\ref{prop:algo} the procedure eliminates $S$.
\end{proof}
The corollary applies to every expression that the procedure meets: the terms of the output of ID
are minimal (Section~\ref{sec:setting}), and along a run they remain so, because \Proc{Factorize}
minimises the terms it returns (Example~\ref{ex:restore}); minimisation does not prevent an
elimination, since a joint assembled from terms with idle conditions is an assembly for the minimal
terms, in which the terms that contain $S$ only as an idle condition count among the retained terms
of Definition~\ref{def:fullreduced}. The corollary carries two qualifications: it concerns a single
elimination, and it is relative to joins and insertions. The first matters because a run is a
sequence of eliminations, whose result can depend on their order, for both procedures alike
(Example~\ref{ex:order}); the end of Section~\ref{sec:algo} describes how several orders could be
explored, and the completeness of the single elimination is not affected.

As for joins and insertions, every part of an elimination that uses them is fixed: the terms to
assemble (Proposition~\ref{prop:reduce}), the set of
each join (Theorem~\ref{thm:join}), the variables to insert and the term inserted for each
(Section~\ref{sec:insert}); what remains open is whether steps of another kind exist to create an assembly. A summation disappears only through (R1), on a joint distribution in which
the summed variable is a head, and joins and insertions are the only steps we know that produce such
a joint from the terms of an atomic expression by the rules (R1)--(R4) and the conditional
independences implied by $G$. We conjecture that no other step of this kind exists; if so, the
procedure is complete among rewritings by conditional independences.\footnote{Rewritings of another kind
do exist, because graphs with latent variables also impose equality constraints that are not conditional
independences of $P$, such as the Verma constraint \cite{vermapearl1990}, which the nested Markov
model characterises as independences in distributions obtained from $P$ by fixing variables
\cite{richardsonetal2023}. However, the authors are not aware of a use of such constraints to remove summations}

\section{Conclusion}
\label{sec:concl}
 
The only rule by which a summation disappears is (R1): a conditional distribution sums to one over
its head. Removing a summation therefore amounts to assembling terms into a joint distribution in
which the summed variable is a head, and TK builds it by joins and by
the insertion of missing variables, an effective device that supplies heads that no term provides.
We have made two changes: only the terms that contain the summed variable are assembled, and every
term is kept minimal, minimality being restored after every summation. Together they fix the rest: the join set, the inserted variables and the inserted terms are each
decided without a search, and topological consistency
is not needed. The resulting procedure is sound, removes summations that the original procedure
leaves in place.
 
Within joins and insertions the procedure is complete for each elimination
(Corollary~\ref{cor:complete}), although the result of a whole run can depend on the order of the
eliminations (Example~\ref{ex:order}); and if, as
we conjecture, no other step based on conditional independences can bring a summation to the form that
(R1) requires, it is complete among such rewritings. Equality constraints that are not independences,
such as the Verma constraint, lie outside this scope (Section~\ref{sec:sc}). Some summations cannot be removed at all, such as the adjustment for $B$ in
Example~\ref{ex:setaside}, which expresses an average over the summed variable; simplifications of
other kinds, such as identities between the numerator and the denominator of a ratio, lie outside
this scope.
 
\section*{Acknowledgements}
 
We thank S. Tikka for early discussions about \cite{tikkakarvanen2017} and E. Armengol for suggesting that two minimal terms are joined at the unique minimal set $\Cc$ of
Theorem~\ref{thm:join}. The graphs and queries presented as examples were found with the help of
the AI model Claude (Anthropic), after dozens of hours of discussion under the guidance of the
authors. The authors are supported by UKRI Future Leaders Fellowship Grant No. 10128920.
 
\appendix
 
\section{Comparison with TK}
\label{app:reduce}
 
This appendix contains the proof of Proposition~\ref{prop:reduce}, by which the terms and variables
that TK assembles besides the participating ones never help an elimination
(Appendix~\ref{app:step1}), and the argument, used after that proposition and in the proof of
Corollary~\ref{cor:complete}, that an assembly can always be rearranged into decreasing topological
order (Appendix~\ref{app:step2}). With Corollary~\ref{cor:complete}, these two results establish
that the procedure of this paper eliminates a summation variable from an expression whenever TK
does. Appendix~\ref{app:step3} then turns to sequences of eliminations, through an example in which
their order changes the result.
 
\subsection{Full and reduced assemblies}\label{app:step1}
Let $P(\Ab\mid\Bb)$ be a full assembly for $S$ and $P(\Hb\cup\Rb\mid\Lb)$ its reduction
(Definition~\ref{def:fullreduced}). The two joints have different chain rules, and a prime marks
the full one. By Definition~\ref{def:assembly}, a head $g$ is conditioned in the full joint on
\[
  \Eb'_g:=\Bb\cup\{x\in\Ab: x\text{ assembled after }g\},
\]
and in the reduced joint on $\Eb_g:=\Lb\cup\{x\in\Hb\cup\Rb: x\text{ assembled after }g\}$, which in
decreasing topological order is the set~\eqref{eq:EM}. The variables of
$\Hb\cup\Nb$ among the conditions of the full joint are those of $\Lb$, and among its heads those of
$\Hb\cup\Rb$, so that
\[
  \Eb_g=\Eb'_g\cap(\Hb\cup\Nb):
\]
the reduction conditions each term on the same variables of $\Hb\cup\Nb$ as the full assembly, and on
none of $\Ub$. In Example~\ref{ex:setaside}, for instance, TK
tries to eliminate $A$ through the full joint $P(Y,B,A\mid X)$, assembled in the order $Y,B,A$, with
$\Hb=\{A,B\}$, $\Nb=\emptyset$ and $\Ub=\{X,Y\}$; the participating term $P(B\mid A)$ must equal
$P(B\mid\Eb'_B)=P(B\mid A,X)$, which fails because of $X\leftrightarrow B$, whereas in the reduction it
must equal only $P(B\mid\Eb_B)=P(B\mid A)$.
 
\begin{proof}[Proof of Proposition~\ref{prop:reduce}]
\emph{The participating terms.} For $h\in\Hb$, $P(h\mid C_h)=P(h\mid\Eb'_h)$, since the
term is the factor of $h$ in the full joint, and $C_h\subseteq\Eb'_h\cap(\Hb\cup\Nb)=\Eb_h$. The term does
not depend on the conditions in $\Eb'_h\setminus C_h$, hence, by decomposition, not on those in the
subset $\Eb_h\setminus C_h$, and $P(h\mid C_h)=P(h\mid\Eb_h)$. By the chain rule, the participating
terms and the terms $P(M\mid\Eb_M)$, $M\in\Rb$, multiply to $P(\Hb\cup\Rb\mid\Lb)$.
 
\emph{The inserted terms.} The full assembly uses for $M\in\Rb$ the term $P(M\mid\Eb'_M)$, free of
$S$; the reduction inserts $P(M\mid\Eb_M)$, which lacks the conditions in $\Ub$, and dropping a
condition may create a dependence on $S$. As in Corollary~\ref{cor:complete}, we compare products
rather than single terms. Let $Q$ be the product of the terms of the full assembly other than the
participating ones, which is free of $S$, so that $P(\Ab\mid\Bb)=\prod_{h\in\Hb}P(h\mid C_h)\,Q$.
Multiplying by $P(\Bb)$ and summing over the variables of $\Ub$, which occur in no participating term
and do not include $S$, gives $P(\Hb\cup\Nb)=\prod_{h\in\Hb}P(h\mid C_h)\sum_{\Ub}P(\Bb)\,Q$. Since
$P(\Hb\cup\Rb\mid\Lb)=P(\Hb\cup\Nb)/P(\Lb)$, the first part gives
\[
  \prod_{M\in\Rb}P(M\mid\Eb_M)=\frac{P(\Hb\cup\Rb\mid\Lb)}{\prod_{h\in\Hb}P(h\mid C_h)}
  =\frac{1}{P(\Lb)}\sum_{\Ub}P(\Bb)\,Q,
\]
which is free of $S$, as none of $P(\Lb)$, $P(\Bb)$ and $Q$ contains it. Each factor is then free of
$S$: the variable of $\Rb$ assembled first occurs only in its own factor, so summing over it shows
that the other factors multiply to a function free of $S$, and the argument proceeds down the order. The reduction is therefore an assembly for $S$, whose participating terms are the factors of their heads and whose inserted terms are free of $S$.
\end{proof}
 
\subsection{The order of assembly}\label{app:step2}
Definition~\ref{def:assembly} takes the heads in decreasing topological order. A joint assembled in
another order, such as the one that TK builds in Example~\ref{ex:order}, can be rearranged into this
order without changing its terms. Let $u$ be assembled immediately before $w$, with $u<w$, and let
$\mathbf{Z}$ consist of the conditions of the joint and of the heads assembled after $w$. The chain
rule gives $u$ the term $P(u\mid\mathbf{Z},w)$ and $w$ the term $P(w\mid\mathbf{Z})$. The term of $u$
depends only on variables below $u$, hence not on $w$, so that $u\indep w\mid\mathbf{Z}$; then
$P(w\mid\mathbf{Z},u)=P(w\mid\mathbf{Z})$ and $P(u\mid\mathbf{Z})=P(u\mid\mathbf{Z},w)$. Assembling
$w$ before $u$ therefore leaves both terms unchanged, and exchanging in this way every adjacent pair
in the wrong order brings the joint into decreasing topological order, with the same inserted terms.
The argument fails for an inserted term that depends on a variable above its head, which could not
be moved; TK, however, could not divide out such a term either, since its factorisation writes the
result by the chain rule in topological order.
 
\subsection{Sequences of eliminations}\label{app:step3}
The results above compare the two procedures on one elimination, from one expression, whereas a
run is a sequence of eliminations, each applied to the expression that the previous one has
written. Sequences deserve a study of their own, which we do not attempt here: an elimination can
prevent a later one, for both procedures alike, so that the result of a run depends on the order in
which the summation variables are tried, and the order followed by the procedure of this paper,
from the highest summation variable down, is not always the best one. In the following example the
procedure of this paper, tried in its default order, ends with two summations where TK ends with
one, because it eliminates a variable that TK cannot eliminate, whereas in the order that TK ends
up following it reaches the same expression as TK.

\begin{example}[the order of the eliminations matters]\label{ex:order}
Consider the graph below, with the topological order $v_0<v_1<\dots<v_6$, and the query
$P(v_3,v_6\mid \hat{v}_5)$.
\begin{center}
\begin{tikzpicture}[node distance=12mm,yscale=0.75]
\node[vtx] (v0) at (0,0) {$v_0$};
\node[vtx] (v1) at (0,-1.8) {$v_1$};
\node[vtx] (v2) at (2,0.9) {$v_2$};
\node[vtx] (v3) at (2,-0.9) {$v_3$};
\node[vtx] (v4) at (4,-1.8) {$v_4$};
\node[vtx] (v5) at (4,0.9) {$v_5$};
\node[vtx] (v6) at (6,0) {$v_6$};
\draw[dir] (v0) -- (v2);
\draw[dir] (v0) -- (v3);
\draw[dir] (v0) to[bend left=55] (v6);
\draw[dir] (v1) -- (v4);
\draw[dir] (v2) -- (v3);
\draw[dir] (v2) -- (v5);
\draw[dir] (v3) -- (v4);
\draw[dir] (v4) -- (v6);
\draw[dir] (v5) -- (v6);
\draw[bid] (v0) -- (v1);
\draw[bid] (v0) to[bend right=30] (v3);
\draw[bid] (v1) -- (v3);
\draw[bid] (v2) to[bend left=35] (v3);
\draw[bid] (v4) to[bend right=35] (v6);
\end{tikzpicture}
\end{center}
 
ID returns
\[
  \sum_{v_0,v_1,v_2,v_4}P(v_0)\,P(v_1\mid v_0)\,P(v_2\mid v_0)\,P(v_3\mid v_0,v_1,v_2)\,
  P(v_4\mid v_1,v_3)\,P(v_6\mid v_0,v_1,v_3,v_4,v_5).
\]
 
The procedure of this paper, \Proc{Simplify} (Algorithm~\ref{alg:simplify}), tries $v_4$ first.
Its participating terms are $P(v_4\mid v_1,v_3)$ and $P(v_6\mid v_0,v_1,v_3,v_4,v_5)$, so that
$\Hb=\{v_4,v_6\}$, $\Nb=\{v_0,v_1,v_3,v_5\}$, $\Rb=\{v_5\}$ and $\Lb=\{v_0,v_1,v_3\}$, and the loop runs
through $v_6,v_5,v_4$. The term for $v_6$ is the initial joint; \Proc{Insert} computes
$\Eb_{v_5}=\{v_0,v_1,v_3,v_4\}$, finds that condition~(b), $v_5\indep v_4\mid v_0,v_1,v_3$, holds, and
joins the inserted term $P(v_5\mid v_0,v_1,v_3)$. \Proc{Join} then joins
$v_4$ at $\{v_0,v_1,v_3\}$, since $v_4\indep v_0\mid v_1,v_3$, and \Proc{Factorize} sums out $v_4$
and divides by the inserted term:
\begin{equation}\label{eq:exA}
  \sum_{v_0,v_1,v_2}P(v_0)\,P(v_1\mid v_0)\,P(v_2\mid v_0)\,P(v_3\mid v_0,v_1,v_2)\,
  P(v_6\mid v_0,v_1,v_3,v_5).
\end{equation}
The procedure starts again from the highest summation variable, $v_2$. Its participating terms are
$P(v_2\mid v_0)$ and $P(v_3\mid v_0,v_1,v_2)$; the term for $v_6$ does not contain $v_2$ and is not
assembled. Here $\Rb=\emptyset$, and $v_2$ is joined to $v_3$ at $\{v_0,v_1\}$, since
$v_2\indep v_1\mid v_0$, which gives
\begin{equation}\label{eq:exB}
  \sum_{v_0,v_1}P(v_0)\,P(v_1\mid v_0)\,P(v_3\mid v_0,v_1)\,P(v_6\mid v_0,v_1,v_3,v_5).
\end{equation}
From here neither $v_1$ nor $v_0$ can be eliminated, because each needs the insertion of $v_5$.
For $v_1$, $\Hb=\{v_1,v_3,v_6\}$, $\Rb=\{v_5\}$ and
$\Lb=\{v_0\}$, and this insertion, with $\Eb_{v_5}=\{v_0,v_1,v_3\}$, requires
$v_5\indep v_1\mid v_0,v_3$, which fails on the path $v_5\leftarrow v_2\leftrightarrow v_3\leftrightarrow
v_1$, where $v_3$ is a conditioned collider and $v_2$ is not given. For $v_0$ the insertion of $v_5$
requires $v_5\indep v_0\mid v_1,v_3$, which fails on $v_5\leftarrow v_2\leftarrow v_0$, and the procedure
stops at \eqref{eq:exB}, with two summations.
 
Suppose instead that, after $v_4$, the procedure tries $v_1$ on \eqref{eq:exA}.
Since $v_2$ still occurs, now $\Nb=\{v_0,v_2,v_5\}$, $\Rb=\{v_2,v_5\}$ and $\Lb=\{v_0\}$, and the loop
runs through $v_6,v_5,v_3,v_2,v_1$. The insertion of $v_5$, with $\Eb_{v_5}=\{v_0,v_1,v_2,v_3\}$,
requires $v_5\indep v_1\mid v_0,v_2,v_3$, which holds because $v_2$ blocks the path above, and the
inserted term is $P(v_5\mid v_0,v_2,v_3)$, which equals $P(v_5\mid v_2)$. Then $v_3$ is joined, $v_2$ is inserted with
$P(v_2\mid v_0)$, since $v_2\indep v_1\mid v_0$, and $v_1$ is joined at $\{v_0\}$, which gives
\begin{equation}\label{eq:exD}
  \sum_{v_0,v_2}P(v_0)\,P(v_2\mid v_0)\,P(v_3\mid v_0,v_2)\,P(v_6\mid v_0,v_2,v_3,v_5).
\end{equation}
In the same way $v_0$ is eliminated, inserting $v_5$ with $P(v_5\mid v_2,v_3)=P(v_5\mid v_2)$, since
$v_5\indep v_0\mid v_2,v_3$, and the result is
\begin{equation}\label{eq:exC}
  \sum_{v_2}P(v_2)\,P(v_3\mid v_2)\,P(v_6\mid v_2,v_3,v_5),
\end{equation}
with one summation instead of two; $v_2$ can no longer be eliminated, since $v_5\indep v_2\mid v_3$
fails. Both \eqref{eq:exB} and \eqref{eq:exC} are correct, and neither can be simplified further by
joins and insertions. A summation variable above $S$ can thus be needed, in the conditioning set of an
inserted term, for the elimination of $S$.
 
TK, recalled in Section~\ref{sec:setting}, assembles for each summation variable $S$ the
full joint of the terms whose heads lie between $S$ and the highest head
(Definition~\ref{def:fullreduced}), from the top down, inserts a missing variable when a join
fails, and does not minimise.
 
For $v_4$ it joins $v_4$ to $v_6$ at $\{v_0,v_1,v_2,v_3,v_5\}$: $v_2$ enters as an ancestor of
$v_4$, and remains, idle, in the new term $P(v_6\mid v_0,v_1,v_2,v_3,v_5)$.
 
For $v_2$ it fails, because the full assembly contains, besides the terms for $v_3$ and $v_2$, the term for
$v_6$, which the reduced assembly leaves out and which depends on $v_5$; the join of $v_3$ then keeps $v_5$ among
the conditions of the joint. A candidate set for the join of $v_2$ either drops $v_5$, on which the
joint depends, or keeps it, and then requires $v_2$ to be independent of $v_5$ given $v_0$; the
insertion of $v_5$ requires $v_5$ to be independent of $v_2$ given $v_0$ and possibly $v_1$; both
are excluded by the edge $v_2\to v_5$. The idle condition plays no part: the failure occurs as well when the term for $v_6$ is
minimal.
 
For $v_1$ it succeeds, in a way that the procedure of this paper never follows. After the join of $v_3$ at
$\{v_0,v_1,v_2,v_5\}$, the joint $P(v_6,v_3\mid v_0,v_1,v_2,v_5)$ conditions $v_3$ on $v_5$, which lies
above it. The join of $v_2$ fails, and $v_5$, a condition of the joint, is inserted at
$\{v_0,v_1,v_2\}$, after $v_3$, that is, out of topological order; then $v_2$ is joined at
$\{v_0,v_1\}$ and $v_1$ at $\{v_0\}$, and the result is \eqref{eq:exD}. The procedure of this paper inserts
$v_5$ before $v_3$, since its loop runs from the top, and never conditions a head on a variable above
it. The two orders are equivalent here, because the join of $v_3$ has checked that $v_3\indep v_5\mid v_0,v_1,v_2$, so that the
condition $v_5$ is idle, and by Appendix~\ref{app:step2} the two orders of assembly give the same terms. After another
failed attempt on $v_2$, TK eliminates $v_0$ in the same way and
ends at \eqref{eq:exC}.
 
The greater power of the reduced assembly is thus an obstacle in this example: by removing $v_2$,
which TK cannot remove, the procedure of this paper loses the
variable that the later insertions of $v_5$ need, while TK, held up at $v_2$, keeps it.
As shown above, the procedure of this paper reaches \eqref{eq:exC} when it follows the same
sequence, and a search over orders, as described at the end of Section~\ref{sec:algo}, finds it.
\end{example}


\begin{thebibliography}{99}
 
\bibitem{pearl1995}
J.~Pearl.
\newblock Causal diagrams for empirical research.
\newblock \emph{Biometrika}, 82(4):669--710, 1995.
 
\bibitem{pearl2009}
J.~Pearl.
\newblock \emph{Causality: Models, Reasoning, and Inference}.
\newblock Cambridge University Press, 2nd edition, 2009.
 
\bibitem{shpitserpearl2006a}
I.~Shpitser and J.~Pearl.
\newblock Identification of joint interventional distributions in recursive semi-Markovian
causal models.
\newblock In \emph{Proceedings of the 21st National Conference on Artificial Intelligence},
pages 1219--1226, 2006.
 
\bibitem{huangvaltorta2006}
Y.~Huang and M.~Valtorta.
\newblock Pearl's calculus of intervention is complete.
\newblock In \emph{Proceedings of the 22nd Conference on Uncertainty in Artificial
Intelligence}, pages 217--224, 2006.
 
\bibitem{tianpearl2002}
J.~Tian and J.~Pearl.
\newblock A general identification condition for causal effects.
\newblock In \emph{Proceedings of the 18th National Conference on Artificial Intelligence},
pages 567--573, 2002.
 
\bibitem{tianpearl2003}
J.~Tian and J.~Pearl.
\newblock On the identification of causal effects.
\newblock Technical Report R-290-L, Cognitive Systems Laboratory, University of California, Los
Angeles, 2003.
\bibitem{Lodato:25}
I. Lodato and A. V. Iyer and I. Z. To.
\newblock Bounding interventional queries from generalized incomplete contingency tables.
\newblock https://arxiv.org/abs/2511.05755, 2025.

 
\bibitem{tikkakarvanen2017}
S.~Tikka and J.~Karvanen.
\newblock Simplifying probabilistic expressions in causal inference.
\newblock \emph{Journal of Machine Learning Research}, 18(36):1--30, 2017.
 
\bibitem{geigervermapearl1990}
D.~Geiger, T.~Verma and J.~Pearl.
\newblock Identifying independence in {B}ayesian networks.
\newblock \emph{Networks}, 20(5):507--534, 1990.
 
\bibitem{vermapearl1990}
T.~Verma and J.~Pearl.
\newblock Equivalence and synthesis of causal models.
\newblock In \emph{Proceedings of the 6th Conference on Uncertainty in Artificial Intelligence},
  pages 255--270, 1990.
 
\bibitem{richardsonetal2023}
T.~S. Richardson, R.~J. Evans, J.~M. Robins and I.~Shpitser.
\newblock Nested {M}arkov properties for acyclic directed mixed graphs.
\newblock \emph{The Annals of Statistics}, 51(1):334--361, 2023.
 
\end{thebibliography}
\end{document}